\documentclass[final,5p,times,twocolumn]{elsarticle}

\usepackage{amssymb}
\usepackage{amsthm}

\usepackage{amsmath}
\usepackage{siunitx}
\usepackage{subcaption}

\journal{Acta Astronautica}

\begin{document}

\begin{frontmatter}



\title{Optimisation of VLEO Satellite Geometries for Drag Minimisation and Lifetime Extension}


\author[inst1]{F. Hild}

\affiliation[inst1]{organization={Institute of Space Systems, University of Stuttgart},
            addressline={Pfaffenwaldring 29}, 
            city={Stuttgart},
            postcode={70569},
            country={Germany}}

\author[inst1]{C. Traub}
\author[inst1]{M. Pfeiffer}
\author[inst1]{J. Beyer}
\author[inst1]{S. Fasoulas}

\begin{abstract}

The utilisation of the Very Low Earth Orbit (VLEO) region offers significant application specific, technological, operational, and cost benefits.
However, attaining sustained and economically viable VLEO flight is challenging, primarily due to the significant, barely predictable and dynamically changing drag caused by the residual atmosphere, which leads to a rapid deterioration of any spacecraft’s orbit unless mitigated by a combination of active and passive techniques.
This article addresses one passive method by optimising satellite shapes in order to achieve a minimisation of the atmospheric drag force and thus extension of operational lifetime.
Contrary to previous investigations in the field, a constant internal volume is maintained to account for the placement of satellite instruments and payload inside the structure.
Moreover, the satellite geometry is not varied heuristically but optimised via a numerical 2D profile optimisation specifically developed for this purpose.
From the resulting optimal satellite profiles, 3D satellite bodies are derived, which are then verified via the Direct Simulation Monte Carlo method within the open-source particle code PICLas.
In addition, rather unconventional designs, i.e. ring geometries, which are based on the assumption of fully specular particle reflections, are proposed and assessed.
The optimised satellite geometries offer pure passive lifetime extensions of up to 46 $\%$ compared to a GOCE like reference body, while the above-mentioned ring geometries achieve passive lifetime extensions of more than 3000 $\%$.
Finally, the article presents design recommendations for VLEO satellites in dependence of different surface properties.

\end{abstract}




\end{frontmatter}



\section{Introduction}
In the recent past, there has been a growing interest to operate satellites in the Very Low Earth Orbit (VLEO) region, defined as the entirety of orbits with a mean altitude below $\SI{450}{\kilo\meter}$~\cite{livadiotti}.
Besides the potential benefits the regime offers~\cite{benefits-vleo,benefits-vleo-2}, this growth in interest is largely driven by the very real risk of traditional Low-Earth Orbits (LEO) becoming congested and that a satellite deployment in this region is no longer viable due to an extreme risk of hypervelocity collisions.
The risk of incurring a Kessler syndrome~\cite{Kessler} would be greatly exacerbated should one or more of the currently planned mega-constellations of satellites conceived by private companies such as SpaceX, OneWeb, Amazon or Samsung~\cite{Boley} be realised even in part.
With decay times in VLEO being intrinsically short, the amount of potentially harmful debris existing at any given time can be considered sufficiently low to continue traditionally LEO based satellite services from platforms in VLEO, even following the potential fallout of a Kessler syndrome.

The major challenge for a sustained operation in VLEO to become reality, however, is the atmospheric drag caused by the interaction between the residual atmospheric particles and the satellite's surfaces.
Drag continuously dissipates energy from the system, causing a faster rate of orbit decay and thus limited operation times if no adequate means of compensation is available.
Passively, however, the lifetime of a VLEO satellite can be increased by any reduction in the resulting drag force $F\textsubscript{d}$ acting on the satellite, i.e. by means of satellite design adaptions.

The research objective described in this article is the geometrical optimisation of a cylindrical reference satellite, aiming for a passive minimisation of the atmospheric drag and thus an extension of operational lifetime, while ensuring that the internal volume is kept constant.
The aim is to contribute to the aspiring field of the exploitability of the VLEO regime by providing minimum-drag designs for satellites.

This article draws heavily on the master's thesis of the corresponding author~\cite{hild} and is structured as follows.
In Section~\ref{sec:fundamentals}, the relevant fundamentals of satellite aerodynamics are given before the research approach of the work is presented in Section~\ref{sec:research-approach}.
The numerical 2D optimisation for satellite profiles developed for this investigation is explained and the results are presented in Section~\ref{sec:optimisation}.
For the verification of the results, Direct Simulation Monte Carlo (DSMC) simulations have been performed with the open-source particle code PICLas~\cite{piclas} and summarised in Section~\ref{sec:piclas}, leading to design recommendations for satellites in the VLEO regime. 

\section{Fundamentals of Satellite Aerodynamics} \label{sec:fundamentals}

\subsection{Drag Force}
The drag force $\mathbf{F}\textsubscript{d}$ is a result of the interchange of momentum between the residual atmosphere of the Earth and the spacecraft~\cite{drag-modelling}.
It is linearly dependent on the atmospheric density and thus increases with decreasing altitude.
The orbital energy is reduced by the drag force which leads to a reduction in the semi-major axis $a$ of the spacecraft as well as a reduction of the eccentricity $e$ of the orbit, commonly referred to as circularisation~\cite{vallado}.

Due to the atmospheric drag force $\mathbf{F}\textsubscript{d}$, a satellite of mass $m$ experiences an acceleration $\mathbf{a}\textsubscript{d}$:
\begin{equation}
	\mathbf{a}\textsubscript{d} = \frac{\mathbf{F}\textsubscript{d}}{m} \label{eq:F_d}.
\end{equation}
The latter can be calculated as~\cite{livadiotti}:
\begin{equation}
	\mathbf{a}\textsubscript{d} = - \frac{\rho}{2} \lvert \mathbf{v}\textsubscript{rel} \rvert ^2 \cdot \frac{C\textsubscript{d} A\textsubscript{ref}}{m} \cdot \frac{\mathbf{v}\textsubscript{rel}}{\lvert \mathbf{v}\textsubscript{rel} \rvert} \label{eq:a_d}.
\end{equation}
Here, $\rho$ is the density of the atmosphere at the specific altitude while $\mathbf{v}\textsubscript{rel}$ describes the relative velocity of the spacecraft with regards to the rotating atmosphere of the Earth.
Large uncertainties are contained in the both mentioned parameters due to e.g. the level of solar activity and thermospheric winds.
The drag coefficient $C\textsubscript{d}$, which is a dimensionless quantity describing the satellite's drag in a fluid environment, the reference area $A\textsubscript{ref}$ perpendicular to the incident velocity vector as well as the mass $m$ of the spacecraft can be combined to the ballistic coefficient $\beta$:
\begin{equation}
	\beta = \frac{m}{C\textsubscript{d} A\textsubscript{ref}} \label{eq:beta}.
\end{equation}
It comprises all parameters that are influenced by the satellite design.

The aim is now to design optimised satellite geometries in order to passively increase the operational lifetime.
In order to be able to estimate the effects of design changes quickly and efficiently, the following simplified equation for the orbital lifetime of a spacecraft in a circular orbit $t\textsubscript{L}$ is used~\cite{lifetime}:
\begin{equation}
	t\textsubscript{L} = \frac{\beta H_0}{\rho_0 \sqrt{\mu\textsubscript{E} a}}\cdot \left( 1 - \mathrm{exp} \left( - \frac{h_0}{H_0} \right) \cdot \left( 1 + \frac{h_0}{2a} \right) \right). \label{eq:lifetime}
\end{equation}
$\rho_0$ is the density at the initial altitude $h_0$ while $H_0$ is the atmospheric scale height just below the initial altitude, $a$ is the semi-major axis of the orbit and $\mu\textsubscript{E}$ is the standard gravitational parameter of the Earth.
For the analyses in this investigation, the values for $\rho_0$, $h_0$, and $H_0$ are extracted from tables for the piecewise exponential model according to Ref.~\cite{lifetime}.
In a certain altitude, $h_0, a, H_0$, and $\rho_0$ are constant according to the assumptions, and thus, $\beta$, which represents the properties of the spacecraft, is the only parameter that influences the lifetime of the system.

Even though more accurate methods of estimating the satellite lifetime exist, the presented equation is perfectly adequate for the purposes of this investigation, as the interest goes towards the relative improvements due to the geometry optimisation.
For the same reasons, the environmental conditions are fixed to ensure comparability and to avoid any other influences.

By maximising $\beta$, the atmospheric drag force can be minimised (see Equation~\eqref{eq:a_d}) and a maximised operation time $t\textsubscript{L}$ can be achieved according to Equation~\eqref{eq:lifetime}.
With a constant mass assumption, the maximisation of the ballistic coefficient $\beta$ can be achieved by minimising the product of the drag coefficient and the reference area $C\textsubscript{d} A\textsubscript{ref}$:
\begin{equation}
    t\textsubscript{L} \propto \beta \propto \frac{1}{C\textsubscript{d} A\textsubscript{ref}}.
\end{equation}
Whereas drag reduction is frequently associated with reductions in the drag coefficient, it is rather a reduction in the product of the latter and the satellite's reference area and hence, the aim of this investigation is the minimisation of the product $C\textsubscript{d} A\textsubscript{ref}$ in order to achieve a satellite geometry with maximised lifetime.
An example for this is the design of ESA's GOCE satellite with a drag coefficient as high as $C\textsubscript{d} \approx 3.7$ ~\cite{Koppenwallner} which is nevertheless aerodynamically improved due to a comparably small reference area $A\textsubscript{ref}$.

\subsection{Regime Characterisation}
In VLEO altitudes, the atmosphere is so rarefied that the flow can no longer be considered a continuum, but a free molecular flow instead.
The mean free path $\lambda$ of the particles, which is defined as the mean distance between consecutive collisions between particles, is orders of magnitude greater than the characteristic length of a spacecraft immersed in a flow $l\textsubscript{ref}$~\cite{drag-modelling}.
A free molecular flow is particulate in nature and characterised by the domination of gas-surface interactions, i.e. the interaction of particles with the spacecraft's surface, over inter-particle collisions.
Consequently, incident and reflected particles do not influence each other and thus, the flow itself can be considered collisionless~\cite{livadiotti}.
Thus, it may reasonably be assumed that the incident particle flow is not disturbed by the presence of the body in the flow field.

The molecular speed ratio $s$ is a second important parameter that indicates the behaviour of the flow~\cite{drag-modelling}:
\begin{equation}
	s = \frac{v\textsubscript{rel}}{v\textsubscript{a}}.
\end{equation}
$v\textsubscript{rel}$ denotes the absolute value of the macroscopic flow velocity relative to the satellite whereas $v\textsubscript{a}$ is defined as the thermal velocity or the most probable molecular speed of a gas according to a Maxwell-Boltzmann distribution:
\begin{equation}
	v\textsubscript{a} = \sqrt{\frac{2 R T}{M}}.
\end{equation}
It is a random translational motion and depends on the mean molecular mass $M$ and the temperature $T$ of the gas.
$R$ is the universal gas constant.
A classification regarding the flow behaviour can be introduced by this:
If the molecular speed ratio is high, typically $s > 5$~\cite{s}, and the macroscopic velocity is predominant over the thermal speed of the considered gas, the flow is defined as hyperthermal flow and behaves rather like a collimated beam of particles~\cite{drag-modelling}.
For small values of the molecular speed ratio, typically $s < 5$, the flow can be described as a chaotic drifting Maxwellian flow with a high random thermal motion of the gas constituents which is defined as hypothermal flow.
Consequently, all satellite surfaces may be impinged by gas particles even if they are shaded from the flow.
The thermal velocity of the gas particles increases with altitude and thus, this effect becomes stronger.
However, even for altitudes below $\SI{600}{\kilo\meter}$, where a hyperthermal flow can be assumed~\cite{s}, shaded areas of the satellite may be impacted by particles and drag is generated due to their thermal motion to a limited extent.
In general, the effect highly depends on the environmental conditions that are subjected to great variations e.g. by the solar activity, which is why no strict altitude limit can be defined.

\subsection{Surface properties}
Gas-surface interactions are the interaction of particles with the spacecraft's surface where both energy and momentum are transferred to the body by the impinging gas particles, depending on the surface properties.
Usually, two basic, but simplifying types of reflection are considered~\cite{sentman}:
For a specular reflection, the angle of incidence $\theta\textsubscript{i}$ equals the angle of reflection $\theta\textsubscript{r}$.
For a diffuse reflection, the particle is re-emitted according to a probabilistic velocity and direction distribution and thus the angle of reflection $\theta\textsubscript{r}$ is not related to the angle of incidence $\theta\textsubscript{i}$.
For both processes, the velocity of the reflected particle depends on the amount of energy which is transferred to the surface.
The type of reflection that occurs and the energy accommodation to the wall are described by accommodation coefficients and depend on the surface material and its properties as well as environmental conditions.

The thermal or energy accommodation coefficient $\alpha\textsubscript{T}$ describes the energy exchange between the particles and the surface~\cite{livadiotti} and indicates how closely the kinetic energy of the incoming particle has adjusted to the energy of the surface with respect to the wall temperature.
Under the assumption that the translational, rotational and vibrational energies of the considered particle are all affected to the same degree by the gas-surface interaction, $\alpha\textsubscript{T}$ is defined as:
\begin{equation}
	\alpha\textsubscript{T} = \frac{E\textsubscript{i} - E\textsubscript{r}}{E\textsubscript{i} - E\textsubscript{w}}. \label{eq:alpha}
\end{equation}
$E\textsubscript{i}$ and $E\textsubscript{r}$ are the kinetic energies of the incident and reflected particles while $E\textsubscript{w}$ is the kinetic energy a re-emitted particle would have, if it is reflected with the wall temperature.

The momentum exchange may be described by two momentum coefficients, one for the tangential, $\sigma$, and one for the normal momentum, $\sigma'$, as introduced by Schaaf and Chambre~\cite{schaaf-chambre}:
\begin{align}	
	\sigma &= \frac{\tau\textsubscript{i} - \tau\textsubscript{r}}{\tau\textsubscript{i} - \tau\textsubscript{w}} = \frac{\tau\textsubscript{i} - \tau\textsubscript{r}}{\tau\textsubscript{i}},\label{eq:sigma}\\
	\sigma' &= \frac{p\textsubscript{i} - p\textsubscript{r}}{p\textsubscript{i} - p\textsubscript{w}}.\label{eq:sigma'}
\end{align}
Here, $\tau\textsubscript{i}$ and $p\textsubscript{i}$ are the tangential and normal momentum components carried to the surface by an incident particle while $\tau\textsubscript{r}$ and $p\textsubscript{r}$ are the tangential and normal momentum components carried away from the surface by a reflected particle.
$\tau\textsubscript{w}$ and $p\textsubscript{w}$ are the tangential and normal momentum components that would be carried away from the surface by a diffusely reflected particle that is in thermal equilibrium with the surface.
The former is per definition equal to zero due to the symmetrical velocity distribution of the diffuse reflection around the surface normal.

The model of Maxwell~\cite{maxwell} provides another possibility to describe the particle reflection.
For this, $g$ is defined as the fraction of molecules that are reflected diffusely and completely thermally accommodated to the surface, while the remainder $(1-g)$ is being reflected specularly.
The behaviour of the satellite surface is thus described by a linear combination of both reflection types.

Typically, almost completely diffuse reflection and almost full thermal accommodation, i.e. $\alpha\textsubscript{T} \approx 1$, is assumed for VLEO altitudes due to the high number density of atomic oxygen as main atmospheric constituent~\cite{gsi-moe}, which leads to deposition and erosion of the surface.

\subsection{Effects of the Reflection Types on the Drag} \label{sec:rings}
\begin{figure}
	\centering
	\includegraphics[width=\linewidth]{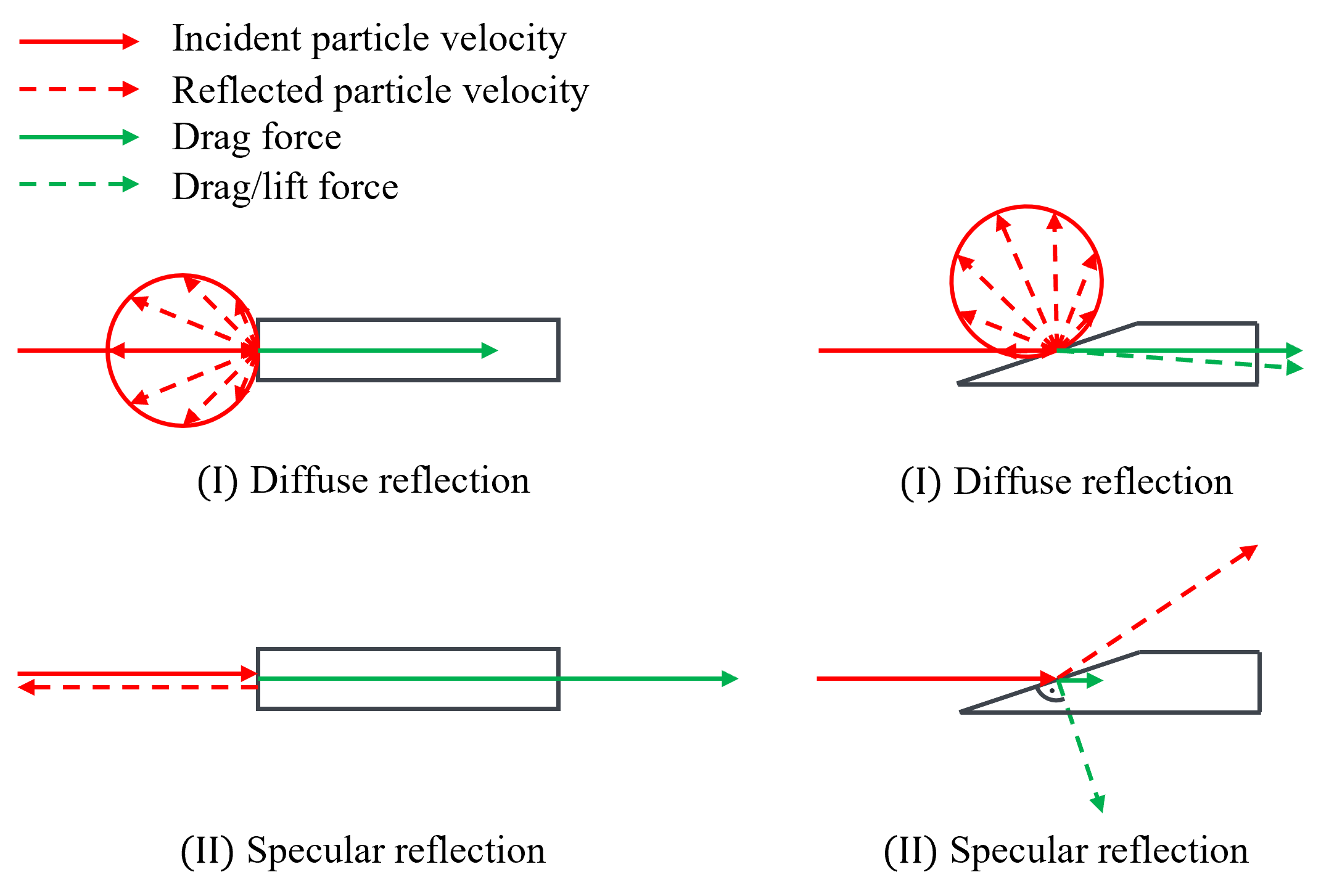}
	\caption{Drag force due to energy and momentum exchange with particles that are (I) diffusely and (II) specularly reflected at the wall for a cylindrical body and a wedge geometry. The velocity of the diffusely reflected particles is overstated. The momentum exchange for the specularly reflected particles is perpendicular to the surface.}
	\label{fig:reflection-geometry-drag}
\end{figure}

The impact on the drag force of the two different particle reflection types for a cylindrical body as well as a wedge geometry are depicted in Figure~\ref{fig:reflection-geometry-drag}.
For the diffuse reflection type, almost no change in the drag force is achieved by geometry variations whereby for specular reflections of the atmospheric particles, the drag highly depends on the body geometry.
Considering small angles between the surface normal and the incident velocity vector, e.g. a cylindrical shape, diffuse particle reflections are preferable, while specular particle reflections are advantageous for geometries with larger angles (e.g. wedge geometries).
The drag can significantly be reduced by this in accordance with the generation of a comparably large lift force.
Considering these results for fully specular reflections of the atmospheric particles, the design of ring geometries offers the conservation of momentum of the incident particles by a reflection in the direction of the incident velocity vector.
The drag force can thus be minimised.

\begin{figure}
	\centering
	\includegraphics[width=\linewidth]{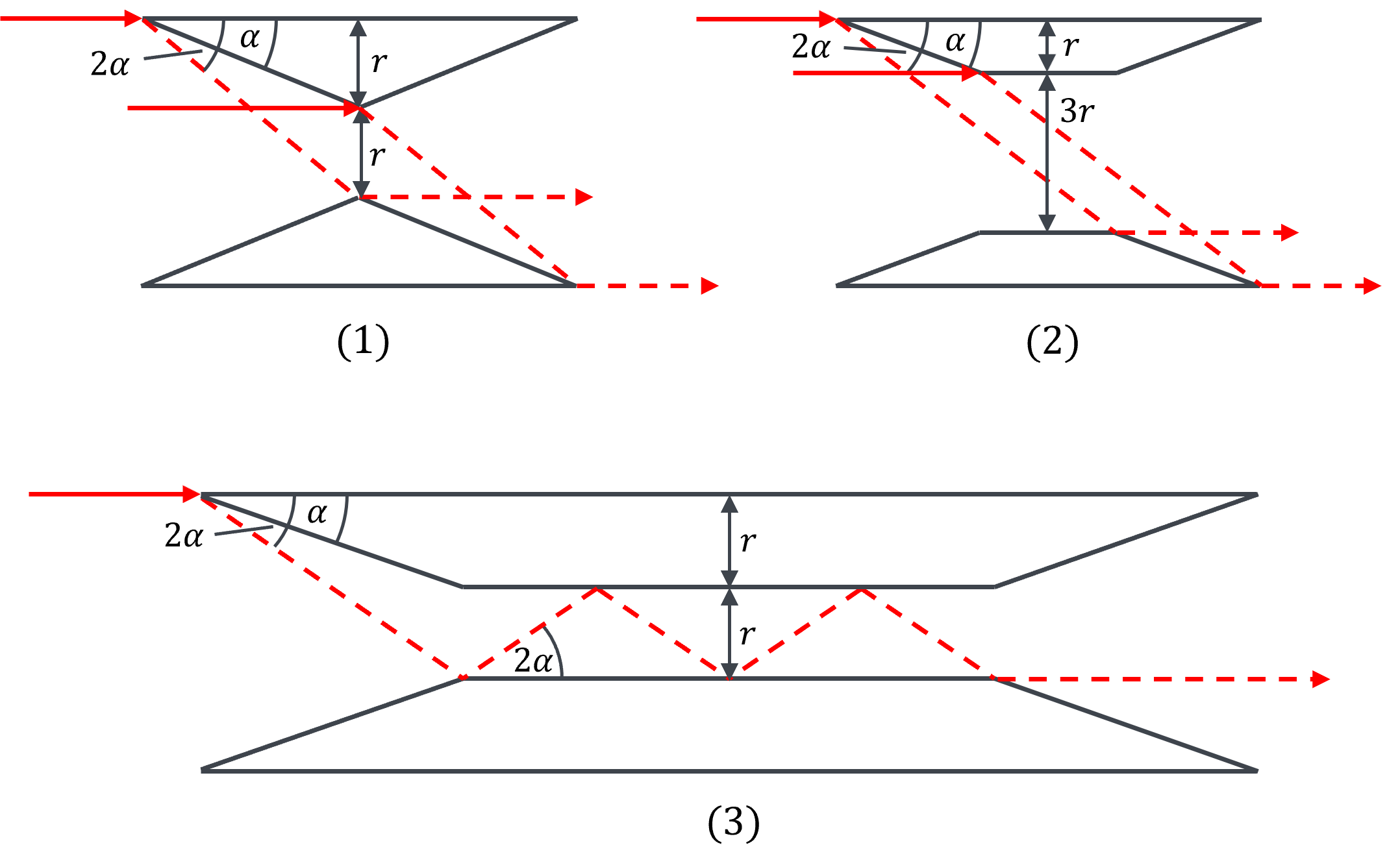}
	\caption{The three considered ring configurations with (1) two specular particle reflections, (2) two specular particle reflections including a tubular middle section and (3) multiple specular particle reflections in an elongated tubular middle section. Particles reflections are indicated with dashed red lines.}
	\label{fig:rings}
\end{figure}

Three different configurations are considered in the further investigations (see Figure~\ref{fig:rings}):
Geometries (1) and (2) include two specular reflections of the particles while design (3) is more flexible in size regarding length and volume restrictions due to the longer tubular section which encourages multiple reflections.

\subsection{State-of-the-Art of Satellite Shape Optimisation}
The state-of-the-art in the field of spacecraft shape design is described in Refs.~\cite{literature-1,literature-3,literature-2,literature-4}.
In the available literature, the satellite shape variations are solely modifications of the front and rear geometries of the body individually and in combination, whereby the internal volume of the bodies is changed.
An overview is given in Table~\ref{t:literature}.
For the investigations, different methods have been applied with 2D and 3D consideration of the satellite bodies, and also different environmental conditions have been used as boundary conditions.
\begin{table}
	\begin{center}
	\small
		\begin{tabular}{|l||c|c|c|c|} 
			\hline
			& Ref. \cite{literature-1} & Ref. \cite{literature-3} & Ref. \cite{literature-2} & Ref. \cite{literature-4} \\
			\hline \hline
			Nose & Yes & Yes & Yes & Yes \\
			geometry & & & & \\
			variation & & & & \\
			\hline
			Rear & No & Yes & No & Yes \\
			geometry & & & & \\
			variation & & & & \\
			\hline
			Constant & No & No & No & No \\
			volume & & & & \\
			\hline
			Altitude & 200, 300 & 250 & 200 & 200 \\
			/ $\si{\kilo\meter}$ & & & & \\
			\hline
			Year & 2014 & 2017 & 2020 & 2021 \\
			\hline
		\end{tabular}
		\caption{Overview of the available literature in the field of spacecraft shape design.}
		\label{t:literature}
	\end{center}
\end{table}

According to these investigations, a reduction in the drag force could be achieved by satellite geometry variations for particle reflections containing a specular component.
Variations in the frontal geometry, i.e. a lowering of the satellite nose angle, are more effective than similar changes in the rear geometry in general due to less interactions between particles and the surface and thus smaller drag generation of areas shaded from the incident flow.
However, the incident flow is not perfectly parallel due to the random thermal motion of the particles and a few particles impact the shaded areas consequently.
The variation of the tail angle can be limited to $8^\circ$ according to Ref.~\cite{literature-4} since there is only limited improvement in drag for larger angles due to a small refill angle of the void behind the body, i.e. the Mach angles of the atmospheric constitutes.
Additionally, Ref.~\cite{literature-4} concludes that there is a greater drag reduction potential for satellites with elongated shapes in general.
For the given boundary and environmental conditions, a maximum reduction in drag of up to $\SI{35}{\percent}$ has been calculated for an elongated body by the variation of the front and rear geometries in combination (see Ref.~\cite{literature-4}).
However, the internal satellite volume is significantly reduced by this and thus the results are hardly qualitatively comparable to the results of the initial body without geometry changes.

\section{Research Approach} \label{sec:research-approach}
\begin{figure}
	\centering
	\includegraphics[width=\linewidth]{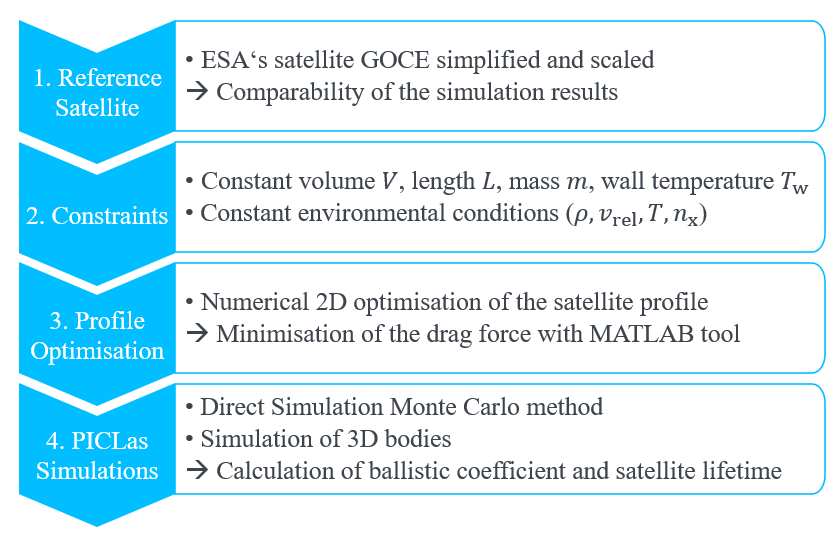}
	\caption{Overview of workflow.}
	\label{fig:workflow}
\end{figure}

In this section, the research approach is presented.
An overview of the overall procedure is illustrated in Figure~\ref{fig:workflow}.

\begin{figure}
	\centering
	\includegraphics[width=\linewidth]{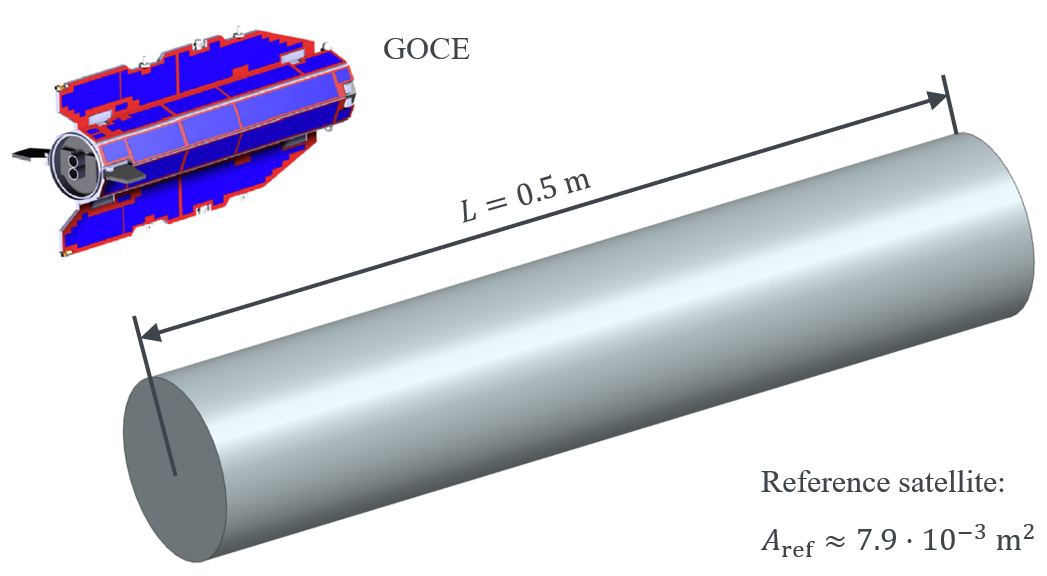}
	\caption{The cylindrical reference body based on a simplification of the ESA's GOCE satellite~\cite{goce-cd}.}
	\label{fig:ref-body}
\end{figure}

\begin{enumerate}
    \item \textit{Reference satellite:} To achieve comparability of the investigation results, a reference body is defined and constraints are fixed first.
    Since it is a prime example for a VLEO orbiting spacecraft, ESA's GOCE satellite~\cite{goce} is selected for this purpose, but simplified and scaled in length to save computing times.
    Its design reflects the current state-of-the-art of satellite shape optimisation for VLEO applications, i.e. a slender satellite shape aiming to minimise the frontal area perpendicular to the flow.
    The elongated cylindrical reference body, excluding solar cells and other booms, is depicted in Figure~\ref{fig:ref-body}.
    \item \textit{Constraints:} Different to previous investigations in the aerodynamic spacecraft design, an internal volume restriction accomplishes a comparability regarding payload and instruments which are to be placed in the satellite.
    The chosen constraint values for the satellite length~$L$, volume~$V$, mass~$m$ and wall temperature~$T\textsubscript{w}$ are listed in Table~\ref{t:satellite}.
    The value for the satellite mass is chosen with $\SI{1}{\kilo\gram}$ per $\si{\cubic\deci\meter}$ since this is typical for a CubeSat.
    The environmental conditions in a circular orbit at the selected altitude of $h = \SI{350}{\kilo\meter}$ are assumed to be constant in addition.
    The relevant parameters according to the NRLMSISE-00 model~\cite{nrlmsise} with moderate solar activity~\cite{iso-solar-indices} are listed in Table~\ref{t:atmosphere-data}.
    The inertial velocity of the circular orbiting spacecraft is used as the relative velocity of the spacecraft to the local flow, i.e. neglecting atmospheric winds and the rotating atmosphere of the Earth.
    \item \textit{2D Profile optimisation:} In contrast to previous investigations, the geometry variations are not only performed heuristically, but via numerical optimisations.
    For this, a MATLAB-based, two-dimensional numerical optimisation of the satellite profile has been developed and implemented with the prescribed constraints and depending on the present surface properties regarding particle reflection types and the energy accommodation of the atmospheric particles to the wall.
    The tool and the optimisation results are explained in greater detail in Section~\ref{sec:optimisation}.
    \item \textit{PICLas simulations:} In order to verify the optimised shapes as well as to account for three-dimensional effects, Direct Simulation Monte Carlo (DSMC) simulations~\cite{bird} of the extended three-dimensional bodies have been conducted and analysed subsequently using the  particle code PICLas~\cite{piclas}.
    The simulation results of all considered satellite designs are analysed to compare the lifetime improvements to the defined reference case.
    This is presented in greater detail in Section~\ref{sec:piclas}.
\end{enumerate}

\begin{table}
	\begin{center}
	\small
		\begin{tabular}{|c l|} 
			\hline
			$L$ & $= \SI{0.5}{\meter}$ \\
			\hline
			$V$ & $= \SI{4e-3}{\cubic\meter}$ \\
			\hline
			$m$ & $= \SI{4}{\kilo\gram}$ \\
			\hline
			$T\textsubscript{w}$ & $= \SI{300}{\kelvin}$ \\
			\hline
		\end{tabular}
		\caption{Satellite body constraints.}
		\label{t:satellite}
	\end{center}
\end{table}
\begin{table}
	\begin{center}
	\small
		\begin{tabular}{|c l|} 
			\hline
			$T$ & $= \SI{1056.6}{\kelvin}$ \\
			\hline
			$\rho$ & $= \SI{9.15e-12}{\kilo\gram\per\cubic\meter}$ \\
			\hline
			$n_\mathrm{O}$ & $= \SI{2.64e14}{\per\cubic\meter}$ \\
			\hline
			$n_{\mathrm{N}\textsubscript{2}}$ & $= \SI{4.18e13}{\per\cubic\meter}$ \\
			\hline
			$n_\mathrm{He}$ & $= \SI{4.88e12}{\per\cubic\meter}$ \\
			\hline
			$n_\mathrm{N}$ & $= \SI{4.44e12}{\per\cubic\meter}$ \\
			\hline
			$n_{\mathrm{O}\textsubscript{2}}$ & $= \SI{1.09e12}{\per\cubic\meter}$ \\
			\hline
			$n_\mathrm{H}$ & $= \SI{8.53e10}{\per\cubic\meter}$ \\
			\hline
			$n_\mathrm{Ar}$ & $= \SI{8.63e9}{\per\cubic\meter}$ \\
			\hline
			$v\textsubscript{rel}$ & $= \SI{7697.1}{\meter\per\second}$ \\
			\hline
			$M$ & $= \SI{0.0174}{\kilo\gram\per\mol}$ \\
			\hline
		\end{tabular}
		\caption{Atmospheric data at $\SI{350}{\kilo\meter}$ altitude according to the NRLMSISE-00 model~\cite{nrlmsise}.}
		\label{t:atmosphere-data}
	\end{center}
\end{table}

\section{2D Profile Optimisation} \label{sec:optimisation}
\subsection{MATLAB Optimisation Tool}
A MATLAB tool for a numerical profile optimisation has been developed in order to minimise the drag solely by changing the geometry of the spacecraft.
The two-dimensional profile of a satellite is defined as the longitudinal section along the direction of the incident velocity vector.
The environmental conditions as well as body volume $V$, length $L$, and wall temperature $T\textsubscript{w}$ are fixed for the optimisation as explained in Section~\ref{sec:research-approach}.
Diverse surface conditions, which are important for the gas-surface interactions, are described by different energy accommodation coefficients $\alpha\textsubscript{T}$ and a varying fraction of diffusely reflected particles $g$.
To take into account the state-of-the-art of surface materials and surface properties in the VLEO region, almost fully diffuse reflection and almost complete thermal accommodation is chosen with $\alpha\textsubscript{T},g = 0.95$ as input parameters for the optimisation.
For possible future improvements of the materials, which promote specular reflections of the atmospheric particles, the parameters $\alpha\textsubscript{T},g = 0.80,\: 0.65$ are considered instead.
By this reduction of the values of $\alpha\textsubscript{T}$ and $g$, it is expected that the drag is reduced by geometry changes according to the analysis of the impact of the reflection types in Section~\ref{sec:rings}.

For the optimisation, the satellite profile is divided into $n$ linear sections in order to precisely determine the drag force acting on each element with Sentman's model~\cite{sentman}.
According to this model, the dimensionless coefficient of the total force component in a particular direction on an area  element assuming free molecular flow is given in~\ref{app:sentman}.
This equation is simplified based on:
\begin{itemize}
    \item the reduction to two dimensions ($\eta=t=0$);
    \item the discretisation of the profile by $n$ linear sections;
    \item the omission of constant parameters like $v\textsubscript{rel}$ and $\rho$;
    \item the summation of the drag forces on the individual profile sections for the calculation of the total force.
\end{itemize}
The optimisation function $D$ with the discretised function values $f_i$ follows:
\begin{equation}
	D = B \cdot f_1 + \sum_{i = 2}^{n+1} \left[ B \sqrt{1 + \left(  \frac{f_i - f_{i-1}}{\Delta x} \right)^2} \Delta x \right],
\end{equation}
with $\Delta x$ being the dimension of each area element in the global $x$ direction and $B$ representing the simplified Equation~\eqref{eq:sentman} from Sentman:
\begin{equation}
    \begin{split}
		B &= \left\{ \left[\sigma \cdot \epsilon k + \left(2 - \sigma' \right) \cdot \gamma l \right] \left[\gamma \left(1 + \mathrm{erf}(\gamma s) \right) + \frac{1}{s \sqrt{\pi}} \mathrm{e}^{-\gamma^2 s^2} \right] \right. \\
		&+ \left. \frac{\left(2 - \sigma' \right) \cdot l}{2 s^2} \left(1 + \mathrm{erf}(\gamma s) \right) \right. \\
		&+ \left. \frac{\sigma' l}{2} \sqrt{\frac{T\textsubscript{w}}{T\textsubscript{i}}} \left[ \frac{\gamma \sqrt{\pi}}{s} \left(1 + \mathrm{erf}(\gamma s) \right) + \frac{1}{s^2} \mathrm{e}^{-\gamma^2 s^2} \right] \right\}.
	\end{split}
\end{equation}
Here, $k,l,t$ are the direction cosines between the direction in which the force is desired and the local $x,y,z$-axes, whereby the $y$-axis is the inward directed surface normal.
Furthermore, $\epsilon, \gamma, \eta$ are the direction cosines between the local $x,y,z$-axes and the macroscopic velocity vector.
Besides this, $T\textsubscript{i}$ is the temperature of the incident particles and $\mathrm{erf}(\gamma s)$ is the error function~\cite{mathematics}.

Subsequently, an optimisation algorithm is applied, whereby the angles of attack of all profile sections are simultaneously and gradually varied to find a local minimum of the function $D$.
This means that a set of all $f_i$ is to be found for which $D$ and thus the drag acting on the profile $\mathrm{d} F / \mathrm{d} z$ becomes minimal.

As Sentman's Equation~\eqref{eq:sentman} depends on the tangential and normal momentum accommodation coefficients $\sigma$ and $\sigma'$, they need to be calculated from the PICLas input parameters $\alpha\textsubscript{T}$ and $g$.
The mathematical correlations are derived in~\ref{app:coefficients}.

Meanwhile, linear and non-linear constraints are considered to account for the maximum dimensions of the body and the prescribed spacecraft volume.
The derivation of the relevant equations is given in~\ref{app:volume}.
By additional linear constraints, the algorithm is further limited to convex profiles to ensure the absence of multiple reflections while it is at the same time only applicable for the absence of surfaces that are shaded from the incident flow to assure the validity of Sentman's model:
\begin{itemize}
    \item A convex surface implies that the slope of a profile section is not greater than the slope of the previous section:
    \begin{equation}
        \forall i: f_{i+1} - f_i \geqq f_{i+2} - f_{i+1}.
    \end{equation}
    \item Non-shaded areas imply that the inclination angle of a section is never negative:
    \begin{equation}
        \forall i: f_{i+1} \geqq f_i.
    \end{equation}
\end{itemize}

Since the numerical optimisation is done in two dimensions, three-dimensional satellite shapes based on the resulting optimal two-dimensional profiles have to be derived for the successive PICLas simulations.
These are referred to as shapes (a), (b) and (c) in the following:
\begin{itemize}
	\item Shape (a): Rotation of the profile along the axis of the incident velocity vector;
	\item Shape (b): Extrusion of the profile with squared shape;
	\item Shape (c): Overlap of the extruded profile and a cylinder with related radius;
\end{itemize}
The different shapes are depicted in Figure \ref{fig:three-shapes} as optimisation results.
Since the volumes of these shapes are calculated in different ways, the choice of a resulting shape is a further input parameter for the optimisation.

Considering additional variations of the tail geometries, the force acting on shaded panels cannot be calculated analytically with Sentman's model.
Therefore, only the front section of the satellite is optimised with the explained optimisation algorithm while the rear part is designed to be a constant flattening profile.

The satellite leaves a void behind it while passing through the fluid~\cite{literature-4}, which slowly refills with atmospheric particles with their respective Mach angle.
The latter is calculated as:
\begin{equation}
	\phi\textsubscript{Refill} = \arctan \left( \frac{\sqrt{\frac{2 R T}{M}}}{v\textsubscript{rel}} \right).
\end{equation}
The refill angles $\phi\textsubscript{Refill}$ for different atmospheric species are listed in Table~\ref{t:refill-angles}.
\begin{table}
	\begin{center}
	\small
		\begin{tabular}{|c|c|} 
			\hline
			Species & $\phi\textsubscript{Refill}$ \\
			\hline \hline
			$\mathrm{O}$ & $7.8^\circ$ \\
			\hline
			$\mathrm{N}\textsubscript{2}$ & $5.9^\circ$ \\
			\hline
			$\mathrm{He}$ & $15.2^\circ$ \\
			\hline
			$\mathrm{N}$ & $8.3^\circ$ \\
			\hline
			$\mathrm{O}\textsubscript{2}$ & $5.5^\circ$ \\
			\hline
			$\mathrm{H}$ & $28.6^\circ$ \\
			\hline
			$\mathrm{Ar}$ & $4.9^\circ$ \\
			\hline
		\end{tabular}
		\caption{Mach angles of the atmospheric species at $\SI{350}{\kilo\meter}$ altitude.}
		\label{t:refill-angles}
	\end{center}
\end{table}
Since the atmosphere mainly consists of atomic oxygen ($\mathrm{O}$) and molecular nitrogen ($\mathrm{N}\textsubscript{2}$) at the considered altitude of $\SI{350}{\kilo\meter}$, an angle of $8^\circ$ is defined as fixed tail angle for all profiles.
By doing so, it can be assumed that only a negligible number of particles impact the shaded surfaces, leading to a reduced drag coefficient of the total body.
The reference area, though, may increase compared to a body without tail variations when considering the volume restriction.
An additional input for the optimisation is the tail length fraction of the total body length, where $25 \: \%$, $50 \: \%$ and $75 \: \%$ have been analysed further within this investigation.

\subsection{Shape Optimisation Results}
\begin{figure}
	\centering
	\includegraphics[width=0.9\linewidth]{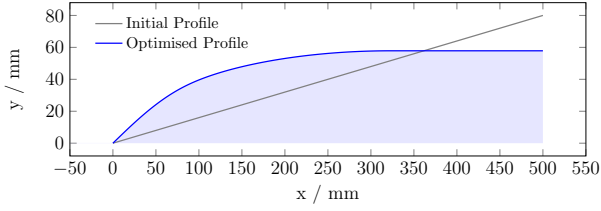}
	\caption{Blue: The optimised profile for $\alpha\textsubscript{T},g = 0.95$ and 3D shape (a). Grey: The initial profile as starting point for the optimisation.}
	\label{fig:matlab-result}
\end{figure}

The numerical optimisation results in a 2D minimum-drag satellite profile depending on the surface conditions with all possible combinations of the energy accommodation coefficient $\alpha\textsubscript{T} = 0.95,\: 0.8,\: 0.65$, the fraction of diffusely reflected particles $g = 0.95,\: 0.8,\: 0.65$, the targeted 3D shape (a), (b), and (c), as well as the tail length fraction of $0 \: \%$, $25 \: \%$, $50 \: \%$, and $75 \: \%$.
The optimised profile for $\alpha\textsubscript{T},g = 0.95$ and 3D shape (a) without rear geometry change is depicted in Figure~\ref{fig:matlab-result} as an example.
The starting values for the optimisation are illustrated in grey.

\begin{figure}
	\centering
	\includegraphics[width=\linewidth]{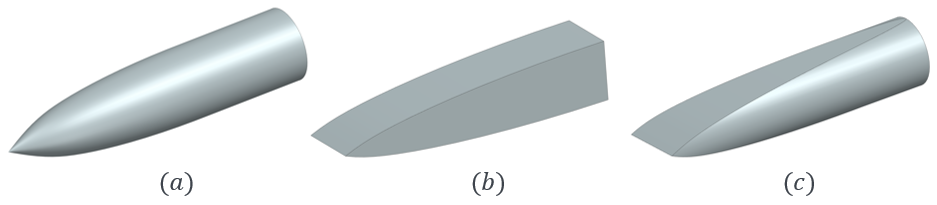}
	\caption{The 3D shapes of the optimised profiles: (a) rotated, (b) extruded, (c) overlap of the both.}
	\label{fig:three-shapes}
\end{figure}

In the following step, each optimised 2D profile is extended to its targeted 3D shape, which is illustrated in Figure~\ref{fig:three-shapes} for all three considered shapes.

\begin{figure}
	\centering
	\includegraphics[width=\linewidth]{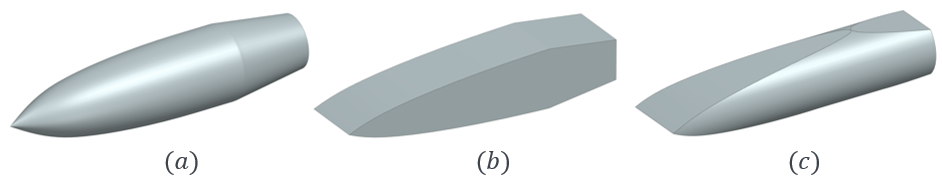}
	\caption{The 3D shapes for the optimised profiles with $25 \: \%$ tail length: (a) rotated, (b) extruded, (c) overlap of the both.}
	\label{fig:three-shapes-tail}
\end{figure}

For additional tail geometry changes, the three available 3D shapes are illustrated in Figure~\ref{fig:three-shapes-tail} for a fraction of $25 \: \%$ of the total tail length as an example.

\section{PICLas Simulations}
\label{sec:piclas}
The resulting 3D bodies from the numerical profile optimisation as well as the ring structures as approach for fully specular particle reflections have been simulated with the DSMC method within the particle code PICLas~\cite{piclas}, which is available on GitHub\footnote{https://github.com/piclas-framework/piclas}.
The results for the drag force and the satellite lifetimes are compared to the prescribed reference case of cylindrical shape and discussed in detail in the following.

\subsection{DSMC and PICLas}
The Direct Simulation Monte Carlo (DSMC) method is a method used for the simulation of rarefied and non-equilibrium flows in general~\cite{bird}. For this purpose, DSMC approximates the Boltzmann Equation~\eqref{eq:boltz} by particles and their motion as well as collision processes, including also relaxation processes of internal degrees of freedom and chemical reactions.
The Boltzmann equation fully describes the behavior of a gas with the corresponding distribution function 
$f=f(\mathbf x, \mathbf v, t)$ at position $\mathbf x$ and velocity $\mathbf v$:
\begin{equation}
    \frac{\partial f}{\partial t} + \mathbf v \frac{\partial f}{\partial \mathbf x} = \left.\frac{\partial f}{\partial t}\right|_{Coll}.
    \label{eq:boltz}
\end{equation}
In this equation, external forces are neglected. Furthermore, $\left.\partial f/\partial t\right|_{Coll}$ is the 
collision term, which can be described by the Boltzmann collision integral:
\begin{equation}
    \left.\frac{\partial f}{\partial t}\right|_{Coll}=\int_{\mathbb{R}^3}\int_{S^2}\mathcal B
    \left[f(\mathbf v')f(\mathbf v_*')-f(\mathbf v)f(\mathbf v_*)\right] \mathrm d\mathbf n\, \mathrm d\mathbf v_*.
\end{equation}
Here, $S^2\subset\mathbb{R}^3$ is the unit sphere, $\mathbf n$ is the unit vector of the scattered velocities, $\mathcal B$ is the collision
kernel and the superscript~$'$ denotes the post collision velocities.

The simulations presented in this article were performed with the DSMC module of the non-equilibrium gas and plasma simulation tool \mbox{PICLas}~\cite{piclas,munz2014coupled}. In addition to the DSMC module, \mbox{PICLas} has the possibility to use other particle methods, e.g. the Particle-In-Cell method (PIC) for treating electromagnetic interactions within the plasma~\cite{birdsall1991particle}. In recent years, further modules have been implemented, such as a particle-based Bhatnagar-Gross-Krook (BGK) operator~\cite{pfeiffer2018particle,pfeiffer2018extending} or a Fokker-Planck operator~\cite{pfeiffer2019evaluation,pfeiffer2017adaptive}, which shall enable more efficient simulations in the transition regime in future work.
However, in this work, only the DSMC module was used.

\subsection{Simulation Cases and Results}
In this work, 2D profiles have been optimised as described in Section~\ref{sec:optimisation} with all possible combinations of the variables displayed in Table~\ref{t:variables}.
The profiles have been extended to the three different described 3D shapes and simulated in PICLas subsequently.
In the following, only the most suitable and promising design results are discussed.
\begin{table}
	\begin{center}
	\small
		\begin{tabular}{|l||l||l|}
			\hline
			Surface properties & 3D shape & Tail geometry \\
			\hline \hline
			\textbullet $\:$ $\alpha\textsubscript{T},g = 0.95$ & \textbullet $\:$ Shape (a) & \textbullet $\:$ No tail \\
			\textbullet $\:$ $\alpha\textsubscript{T},g = 0.80$ & \textbullet $\:$ Shape (b) & \textbullet $\:$ 25 $\%$ of total length \\
			\textbullet $\:$ $\alpha\textsubscript{T},g = 0.65$ & \textbullet $\:$ Shape (c) & \textbullet $\:$ 50 $\%$ of total length \\
			& & \textbullet $\:$ 75 $\%$ of total length \\
			\hline
		\end{tabular}
		\caption{Variables for the profile optimisation and PICLas simulation cases.}
		\label{t:variables}
	\end{center}
\end{table}

\subsubsection{Reference Body}
The drag force acting on the reference body with $\alpha\textsubscript{T},g = 0.95$ is analysed to be $F\textsubscript{d} = \SI{8.11e-06}{\newton}$.
With this result, the drag coefficient is calculated to be $C\textsubscript{d} = 3.81$ according to Equation~\eqref{eq:a_d} and the ballistic coefficient of the body is $\beta = \SI{133.7}{\kilo\gram\per\square\meter}$ utilising Equation~\eqref{eq:beta}.
For the reference satellite under investigation, the lifetime is determined to be 167.1 days.
The results are summarised in Table~\ref{t:reference}.
\begin{table}
	\begin{center}
	\footnotesize
		\begin{tabular}{|c|c|c|c|c|c|} 
			\hline
			$\alpha\textsubscript{T},g$ & $A\textsubscript{ref}$ / $\si{\square\meter}$ & $C\textsubscript{d}$ & $\beta$ / $\si{\kilo\gram\per\square\meter}$ & $t\textsubscript{L}$ / d & $\Delta t\textsubscript{L}$ / $\%$ \\
			\hline \hline
			$0.95$ & & $3.81$ & $133.7$ & $167.1$ & $-$ \\
			\cline{1-1} \cline{3-6}
			$0.80$ & $\num{7.9e-3}$ & $4.01$ & $125.2$ & $156.5$ & $- 6.3$ \\
			\cline{1-1} \cline{3-6}
			$0.65$ & & $4.18$ & $121.8$ & $152.2$ & $- 8.9$ \\
			\hline
		\end{tabular}
		\caption{Simulation results for the reference body.}
		\label{t:reference}
	\end{center}
\end{table}

Due to the fact that the drag force, the inverse of the ballistic coefficient and the satellite lifetime are linear dependent with respect to the underlying assumptions, the following discussion only comprises the lifetime $t\textsubscript{L}$.

\begin{figure}
	\centering
	\includegraphics[width=\linewidth]{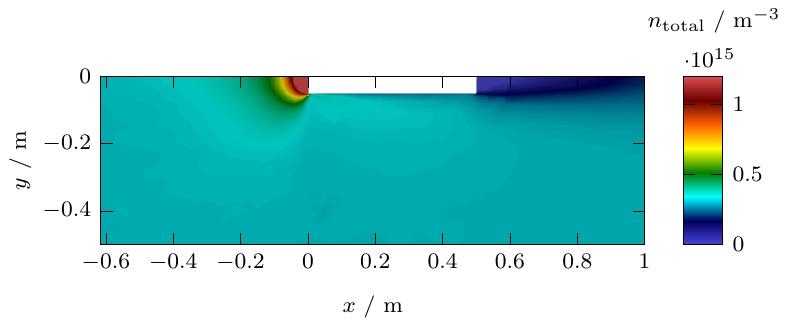}
	\caption{Extract from the simulation result for the total number density of the atmospheric particles in the surrounding area of the reference body for $\alpha\textsubscript{T},g = 0.95$. Only half of the body is depicted in white due to symmetry. The incident velocity is directed in the positive $x$-axis direction.}
	\label{fig:reference-number-density}
\end{figure}

Figure~\ref{fig:reference-number-density} illustrates the simulation result for the particle number density in the surrounding area of the reference body.
It shows the particle accumulation in the area in front of the body while behind the body, a void is visible due to the body movement through the gas.

When further analysing the reference body with changed surface properties, a decreased lifetime by $- \, 6.3 \: \%$ for $\alpha\textsubscript{T},g = 0.80$ and $- \, 8.9 \: \%$ for $\alpha\textsubscript{T},g = 0.65$ is accomplished.
Based on these results, it can be concluded that bodies with a frontal area perpendicular to the incident flow vector are rather suitable for fully diffuse reflection properties of the satellite surfaces.
In the following, all simulation results are compared to the nominal reference case with $\alpha\textsubscript{T},g = 0.95$.

\subsubsection{Optimised Profile Bodies}
The simulation results of the 3D bodies extended from the optimised 2D profiles show that a drag reduction and thus lifetime increase can be achieved by geometry variations.

The best simulation results are summarised in Table~\ref{t:optimised}.
With the rotated body, which is the presented shape (a) (see Figure~\ref{fig:three-shapes}), the lifetime is increased best by $8.9 \: \%$ (+~14.9 days) for $\alpha\textsubscript{T},g = 0.95$ and by $21.1 \: \%$ (+~35.2 days) for $\alpha\textsubscript{T},g = 0.80$, both compared to the reference case.
In contrary, 3D shape (c) shows the highest lifetime increase for $\alpha\textsubscript{T},g = 0.65$ with $+ \, 42.9 \: \%$.
This accomplishes a lifetime extension of 71.7 days compared to the reference case.
\begin{table}
	\begin{center}
	\footnotesize
		\begin{tabular}{|c|c|c|c|c|c|c|} 
			\hline
			$\alpha\textsubscript{T},g$ & Shape & $A\textsubscript{ref}$ / $\si{\square\meter}$ & $C\textsubscript{d}$ & $\beta$ / $\si{\kilo\gram\per\square\meter}$ & $\Delta t\textsubscript{L}$ / $\%$ & $\Delta t\textsubscript{L}$ / d \\
			\hline \hline
			$0.95$ & (a) & $\num{1.05e-2}$ & $2.61$ & $145.6$ & $+ 8.9$ & $+ 14.9$ \\
			\hline
			$0.80$ & (a) & $\num{1.12e-2}$ & $2.20$ & $161.9$ & $+ 21.1$ & $+ 35.2$ \\
			\hline
			$0.65$ & (c) & $\num{1.09e-2}$ & $1.91$ & $191.0$ & $+ 42.9$ & $+ 71.7$ \\
			\hline
		\end{tabular}
		\caption{Best simulation results for optimised bodies compared to the reference simulation.}
		\label{t:optimised}
	\end{center}
\end{table}

\begin{figure}
	\centering
	\includegraphics[width=\linewidth]{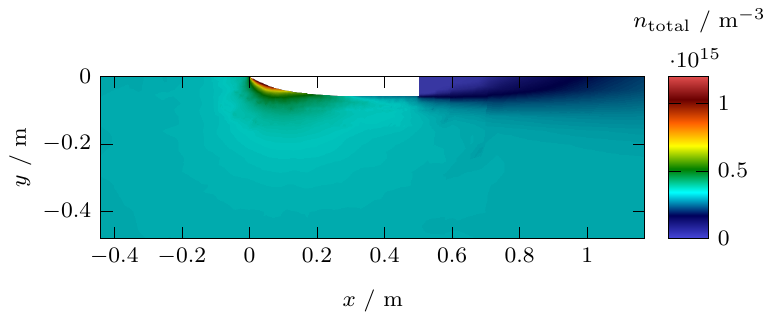}
	\caption{Extract from the simulation result for the total number density of the atmospheric particles in the surrounding area of the optimised rotated profile (shape (a)) for $\alpha\textsubscript{T},g = 0.95$. Only half of the body is depicted in white due to symmetry. The incident velocity is directed in the positive $x$-axis direction.}
	\label{fig:rot-095-number-density}
\end{figure}

In Figure~\ref{fig:rot-095-number-density}, the number density of the atmospheric particles in the simulation domain of the optimised body based on shape (a) with $\alpha\textsubscript{T},g = 0.95$ is shown as an example.
Compared to the reference body in Figure~\ref{fig:reference-number-density}, fewer particles accumulate in the front of the body.

\subsubsection{Optimised Profile Bodies including Tail Variations}
Considering additional variations in the tail geometry with linearly sloping profiles of three different lengths, even greater drag reductions can be accomplished.
The bodies with $25 \: \%$ tail length achieve the best results in lifetime extension for all simulated values of $\alpha\textsubscript{T},g = 0.95, 0.80$, and $0.65$.

In Table~\ref{t:tail-25}, the best simulation results are summarised.
With shape (a), the lifetime can be increased most by $13.0 \: \%$ for $\alpha\textsubscript{T},g = 0.95$ (+~21.7 days) and by $24.2 \: \%$ for $\alpha\textsubscript{T},g = 0.80$ (+~40.5 days).
Similarly, shape (c) achieves the greatest lifetime increase for $\alpha\textsubscript{T},g = 0.65$ with $+ \, 46.3 \: \%$, which corresponds to an extension in lifetime of 77.3 days compared to the reference simulation.
\begin{table}
	\begin{center}
	\footnotesize
		\begin{tabular}{|c|c|c|c|c|c|c|} 
			\hline
			$\alpha\textsubscript{T},g$ & Shape & $A\textsubscript{ref}$ / $\si{\square\meter}$ & $C\textsubscript{d}$ & $\beta$ / $\si{\kilo\gram\per\square\meter}$ & $\Delta t\textsubscript{L}$ / $\%$ & $\Delta t\textsubscript{L}$ / d \\
			\hline \hline
			$0.95$ & (a) & $\num{1.10e-2}$ & $2.40$ & $151.0$ & $+ 13.0$ & $+ 21.7$ \\
			\hline
			$0.80$ & (a) & $\num{1.21e-2}$ & $1.99$ & $166.0$ & $+ 24.2$ & $+ 40.5$ \\
			\hline
			$0.65$ & (c) & $\num{1.06e-2}$ & $1.93$ & $195.5$ & $+ 46.3$ & $+ 77.3$ \\
			\hline
		\end{tabular}
		\caption{Best simulation results for optimised bodies including 25 $\%$ tail length compared to the reference simulation.}
		\label{t:tail-25}
	\end{center}
\end{table}

\begin{figure}
	\centering
	\includegraphics[width=\linewidth]{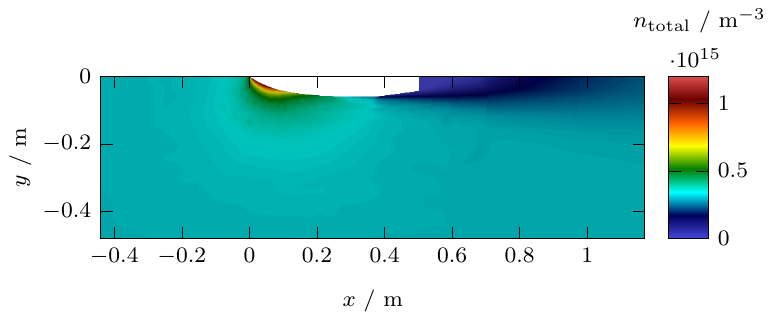}
	\caption{Extract from the simulation result for the total number density of the atmospheric particles in the surrounding area of the optimised rotated profile (shape (a)) with a tail of 25 $\%$ total length for $\alpha\textsubscript{T},g = 0.95$. Only half of the body is depicted in white due to symmetry. The incident velocity is directed in the positive $x$-axis direction.}
	\label{fig:rot-095-tail-25-number-density}
\end{figure}

The particle number density in the simulation domain of the optimised rotated body with 25 $\%$ tail length and $\alpha\textsubscript{T},g = 0.95$ is depicted in Figure~\ref{fig:rot-095-tail-25-number-density}.
Compared to the reference body and the optimised geometry without tail variations, the void behind the body is reduced due to the adaptation of the tail.
It can be seen that the density of the particles is small in the environment of the tail, where the surfaces are shaded from the incident flow.
Thus, the assumption that only few particles impact these surfaces is supported.

For longer tail lengths of 50 $\%$ and 75 $\%$, the reference areas of the optimised bodies increase significantly.
Hence, this compensates any advantage of a decreased value of the drag coefficient $C\textsubscript{d}$ and the drag reduction is considerably smaller than for a tail length of 25 $\%$.
In conclusion, a long tail length is generally not preferable for the design of satellites with a volume restriction.

\begin{figure*}
    \setcounter{subfigure}{0}
	\centering
	\subfloat[No tail.]{
	\includegraphics{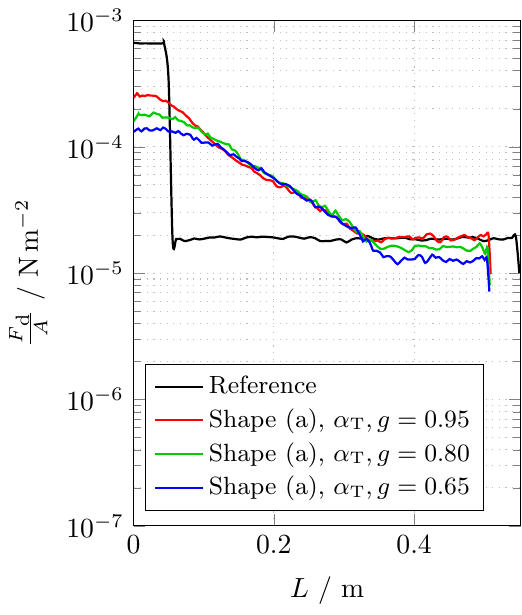}}
	\subfloat[25 $\%$ tail length.]{
	\includegraphics{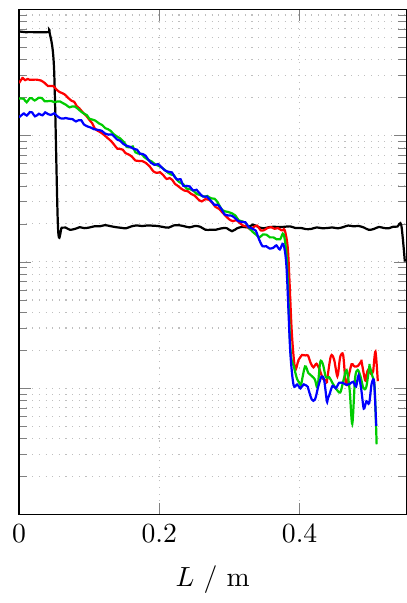}}
	\subfloat[50 $\%$ tail length.]{
	\includegraphics{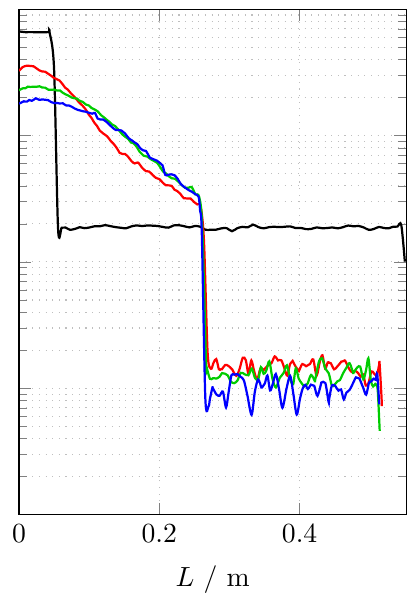}}
	\subfloat[75 $\%$ tail length.]{
	\includegraphics{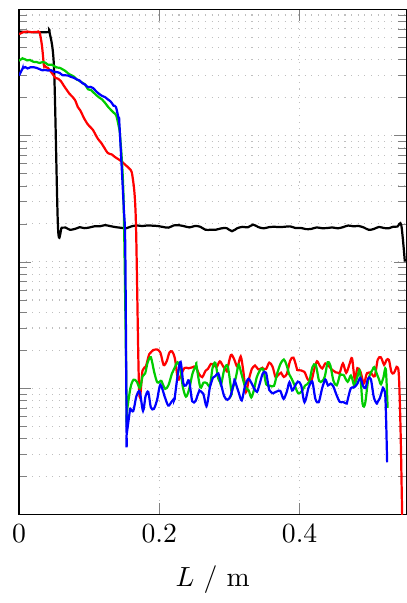}}
	\caption{Simulation results for the drag force per area acting on the optimised rotated profiles, (a) without tail, (b) with $25 \: \%$, (c) $50 \: \%$ and (d) $75 \: \%$ tail length, compared to the reference body.}
	\label{fig:force-rot-tail-no-25-50-75}
\end{figure*}

In Figure~\ref{fig:force-rot-tail-no-25-50-75}, the simulation results of the drag force per area $F\textsubscript{d}/A$ over the body lengths are compared for the different tail lengths.
As an example, the results for the optimised rotated body with shape (a) are illustrated in comparison to the reference body.

While the force per area suddenly decreases due to the sharp edge at the cylinder for the reference body, a slower declination is visible for all optimised bodies due to their slowly decreasing angle of incidence.
Additionally, it can be seen that the force acting on the frontal area of the reference satellite is much higher than the force per area on the optimised bodies.
In general, the force per area is lower for decreasing accommodation coefficients, particularly at the frontal areas.
For the bodies including variations in the rear part, a strong decrease of $F\textsubscript{d}/A$ to values near zero is present due to the small amount of particles impacting the rear surfaces which are shaded from the incident flow.
With an increasing tail length, the force on the front part of the satellite geometries increases while the force is almost equal to zero for the longer tail length region.

\subsubsection{Ring Geometries}
The ring geometries presented in Section~\ref{sec:rings} are based on the assumption of fully specular particle reflections.
Nevertheless, these bodies have been also simulated using PICLas for $\alpha\textsubscript{T},g = 0.95$, 0.80, and 0.65 for comparison.
To account for the fundamental design idea, additional parameter values of $\alpha\textsubscript{T},g = 0.20$ and 0.00 are chosen.

For all considered ring bodies, the lifetime is reduced by up to 33.4 $\%$ for $\alpha\textsubscript{T},g = 0.95$, 0.80, and 0.65 compared to the reference body.
So, these are unsuitable when diffuse particle reflections at the satellite surface are predominant.

The simulation results of the ring geometries for mostly or complete specular particle reflections with $\alpha\textsubscript{T},g = 0.20$ and 0.00 instead promise a significant drag force reduction.
Generally, the lifetime is increasing for decreasing values of $\alpha\textsubscript{T}$ and $g$.
For $\alpha\textsubscript{T},g = 0.20$, a lifetime extension of $+ \, 109.1 \: \%$ is accomplished for the presented ring geometry (2) while for a fully specular reflection of the particles without thermal accommodation, the lifetime is increased by more than $3300 \: \%$.
All simulation results for the ring body (2) are listed in Table~\ref{t:ring}.
Since the ring geometry (2) yield the best results, the other geometries (1) and (3), which have been presented in Section~\ref{sec:rings}, are not further discussed here.
\begin{table}
	\begin{center}
	\footnotesize
		\begin{tabular}{|c|c|c|c|c|c|} 
			\hline
			$\alpha\textsubscript{T},g$ & $A\textsubscript{ref}$ / $\si{\square\meter}$ & $C\textsubscript{d}$ & $\beta$ / $\si{\kilo\gram\per\square\meter}$ & $\Delta t\textsubscript{L}$ / $\%$ & $\Delta t\textsubscript{L}$ / d \\
			\hline \hline
			$0.95$ & & $3.86$ & $91.6$ & $- 31.5$ & $- 52.6$ \\
			\cline{1-1} \cline{3-6}
			$0.80$ & & $3.57$ & $99.0$ & $- 26.0$ & $- 43.4$ \\
			\cline{1-1} \cline{3-6}
			$0.65$ & $\num{1.13e-2}$ & $3.17$ & $111.4$ & $- 16.6$ & $- 27.8$ \\
			\cline{1-1} \cline{3-6}
			$0.20$ & & $1.27$ & $279.5$ & $+ 109.1$ & $+ 182.2$ \\
			\cline{1-1} \cline{3-6}
			$0.00$ & & $0.08$ & $4553.4$ & $+ 3306.3$ & $+ 5524.8$ \\
			\hline
		\end{tabular}
		\caption{Simulation results for ring geometry (2) compared to the reference simulation.}
		\label{t:ring}
	\end{center}
\end{table}

\begin{figure}
	\centering
	\includegraphics[width=\linewidth]{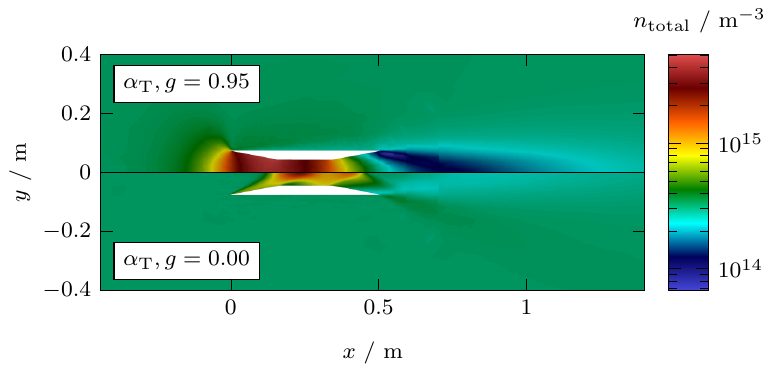}
	\caption{Comparison of an extract from the simulation results for the total number density of the atmospheric particles in the surrounding area of the ring structure (2) for $\alpha\textsubscript{T},g = 0.95$ and $\alpha\textsubscript{T},g = 0.00$. The incident velocity is directed in the positive $x$-axis direction.}
	\label{fig:ring-number-density}
\end{figure}

\begin{figure}
	\centering
	\includegraphics[width=\linewidth]{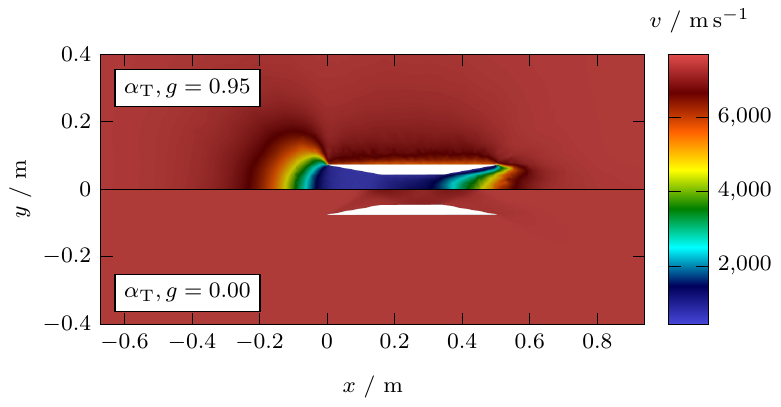}
	\caption{Comparison of an extract from the simulation results for the particle velocity in the surrounding area of the ring structure (2) for $\alpha\textsubscript{T},g = 0.95$ and $\alpha\textsubscript{T},g = 0.00$. The incident velocity is directed in the positive $x$-axis direction.}
	\label{fig:ring-velocity}
\end{figure}

The number density of the atmospheric particles in the simulation domain of ring geometry (2) is displayed in Figure~\ref{fig:ring-number-density} and compared for $\alpha\textsubscript{T},g = 0.95$ and $\alpha\textsubscript{T},g = 0.00$.
Thus, state-of-the-art satellite surfaces in the VLEO region are compared to optimised surface properties, encouraging fully specular reflections.
The number density of the atmospheric particles is significantly lower for the latter with less particle accumulation in the tubular section of the ring.
Less flow rarefaction is present behind the body.
In Figure~\ref{fig:ring-velocity}, the velocities of the atmospheric particles are compared similarly.
For $\alpha\textsubscript{T},g = 0.95$, the particles are decelerated to less than $\SI{500}{\meter\per\second}$ in the tubular section in the middle of the ring whereas for $\alpha\textsubscript{T},g = 0.00$, the particle flow remains almost undisturbed.
In conclusion, the investigated ring geometries accomplish lifetime extensions of up to $+ \, 3300 \: \%$ for materials and surface properties which encourage exclusively specular reflections of the particles.
However, the concept realisation regarding the manufacturing process and the placement of instruments in the ring is certainly challenging.

\subsubsection{Effects of an Angle of Attack}
For the previously presented simulations, the assumption was that the satellites are perfectly aligned with the incident gas flow.
However, an angle of attack to the velocity vector may be present when considering thermospheric winds~\cite{atmos-winds} and uncertainties regarding the spacecraft's attitude in general.
In order to assess the sensitivity of the results for the drag force and lifetime of the satellites discussed so far regarding changes in the satellite's attitude, additional simulations have been conducted with angles of attack of $\chi = 1^\circ$, $2^\circ$, $3^\circ$ and $4^\circ$.
On the basis of the previous results, only the 3D bodies with shape (a) and (c) are considered and the results for $\alpha\textsubscript{T},g = 0.95$ are discussed as an example.

\begin{figure}
	\centering
	\includegraphics[width=0.7\linewidth]{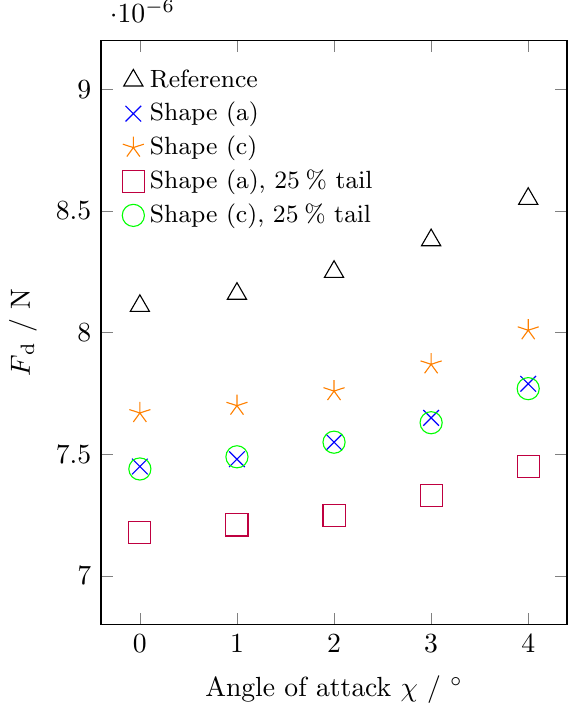}
	\caption{Simulation results for the drag force over the angle of attack $\chi$ for $\alpha\textsubscript{T},g = 0.95$.}
	\label{fig:angle-attack-095}
\end{figure}

In Figure~\ref{fig:angle-attack-095}, the simulation results for the different bodies are illustrated.
In general and as expected, the drag force acting on the bodies increases with the angle of attack $\chi$.
The reference body shows the greatest sensitivity of the drag force with a decrease in the lifetime of $5.1 \: \%$ maximally while the optimised shape (a) with $25 \: \%$ tail length displays the lowest variation with a maximum reduction in $t\textsubscript{L}$ of $3.5 \: \%$, both compared to the respective non-deflected cases.
Even for $\chi = 4^\circ$, the optimised geometries yield better results for the lifetime than the reference body aligned with the incident flow.

\begin{figure}
	\centering
	\includegraphics[width=\linewidth]{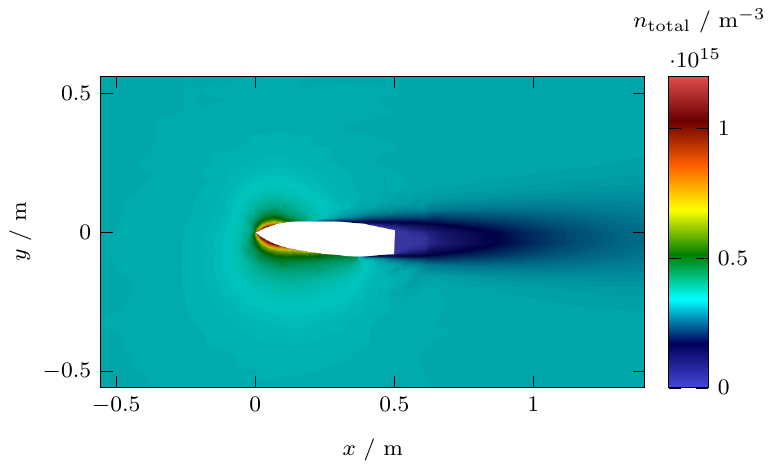}
	\caption{Extract from the simulation result for the total number density of the atmospheric particles in the surrounding area of the optimised rotated body (shape (a)) with a tail length of 25~$\%$ of the total length for $\alpha\textsubscript{T},g = 0.95$ and with an angle of attack of $\chi = 4^\circ$. The incident velocity is directed in the positive $x$-axis direction.}
	\label{fig:angle-attack-rot-095-tail-25-number-density}
\end{figure}

Due to the deflection, the number density of the atmospheric particles below the bodies is greater than above.
This is depicted in Figure~\ref{fig:angle-attack-rot-095-tail-25-number-density} for the optimised rotated body (shape (a)) with 25 $\%$ tail length for $\alpha\textsubscript{T},g = 0.95$ as an example.
By this, an asymmetric force and thus an additional lift force acts on the satellite body.
Whereas frequently been categorised as negligible, this force can eventually be exploited for attitude and relative motion control~\cite{lift}.

\subsection{Design Recommendations}
\begin{figure}
	\centering
	\subfloat[3D satellite geometry.]{\includegraphics[width=0.6\linewidth]{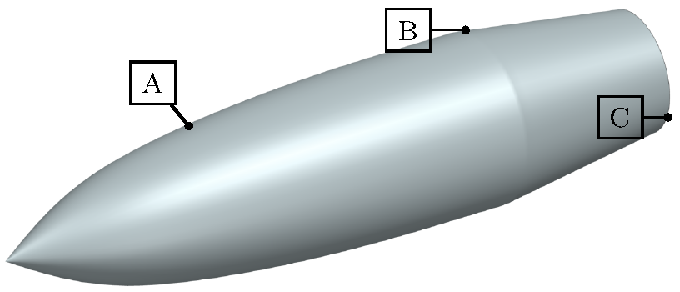}}\\
	\subfloat[Satellite profile (blue). Point A is the intersection point with the reference satellite (grey), point B shows the profile maximum and C denotes the ending point. ]{\includegraphics[width=\linewidth]{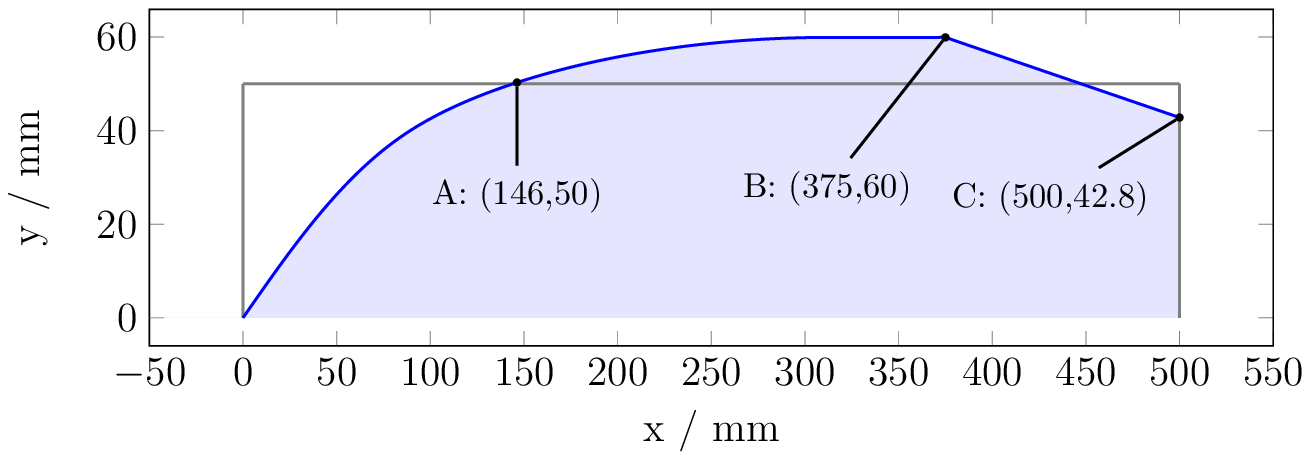}}
	\caption{Optimal satellite geometry for $\alpha\textsubscript{T},g = 0.95$ with shape (a) and 25 $\%$ tail length, achieving $13.0 \: \%$ drag reduction and lifetime extension.}
	\label{fig:optimum-095}
\end{figure}

For state-of-the-art surface properties in VLEO altitudes with the assumption of almost fully diffuse and almost completely thermally accommodated particle reflections ($\alpha\textsubscript{T},g = 0.95$), the best result is achieved by the rotation of the appropriate optimised profile (shape (a)) with a tail fraction of $25 \: \%$, depicted in Figure~\ref{fig:optimum-095}.
A drag reduction of $13.0 \: \%$ is accomplished, leading to a lifetime extension of 21.7 days compared to the reference body with a lifetime of 167.1 days. A greater lifetime extension was accomplished by the solutions suggested in Ref.~\cite{literature-4}, where a maximum drag reduction of $34.6 \:\%$ was achieved due to geometry changes.
However, a internal volume loss of 50 $\%$ was accompanied thereby, where in contrast, for this work, a constant volume is given, making the results hardly comparable.
Additionally, other environmental conditions at an altitude of $\SI{200}{\kilo\meter}$ during high solar activity have been assumed in Ref.~\cite{literature-4}.

\begin{figure}
	\centering
	\subfloat[3D satellite geometry.]{\includegraphics[width=0.6\linewidth]{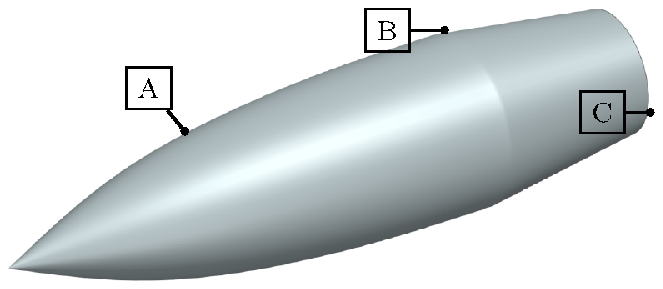}}\\
	\subfloat[Satellite profile (blue). Point A is the intersection point with the reference satellite (grey), point B shows the profile maximum and C denotes the ending point.]{\includegraphics[width=1\linewidth]{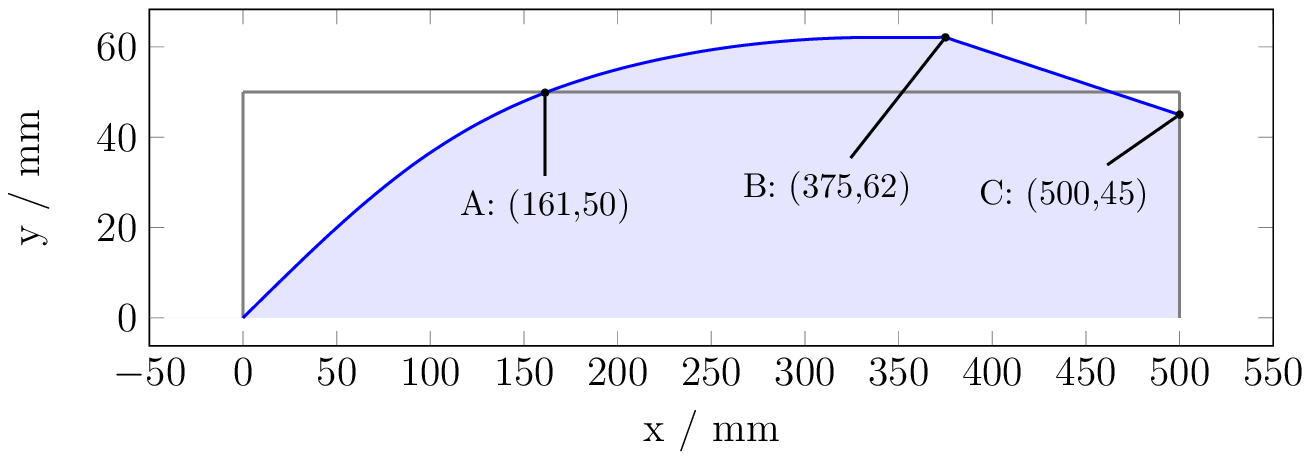}}
	\caption{Optimal satellite geometry for $\alpha\textsubscript{T},g = 0.80$ with shape (a) and 25 $\%$ tail length, achieving $24.2 \: \%$ drag reduction and lifetime extension.}
	\label{fig:optimum-08}
\end{figure}

For a change of the satellite materials and surface properties with the assumption of $\alpha\textsubscript{T},g = 0.80$, the lowest drag is likewise achieved by the optimised satellite profile with 25 $\%$ tail length extended with shape (a), displayed in Figure~\ref{fig:optimum-08}.
The lifetime is increased by $24.2 \: \%$ compared to the reference body, which corresponds to an extension of 40.5 days.

\begin{figure}
	\centering
	\subfloat[3D satellite geometry.]{\includegraphics[width=0.6\linewidth]{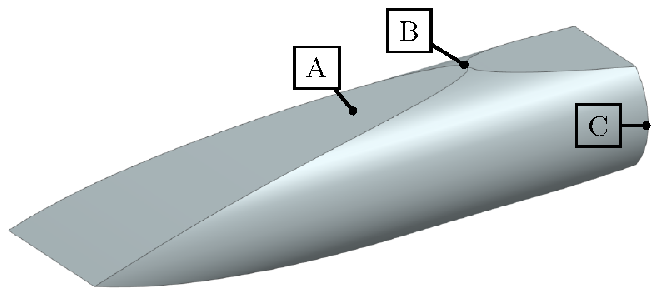}}\\
	\subfloat[Satellite profile (blue). Point A is the intersection point with the reference satellite (grey), point B shows the profile maximum and C denotes the ending point.]{\includegraphics[width=1\linewidth]{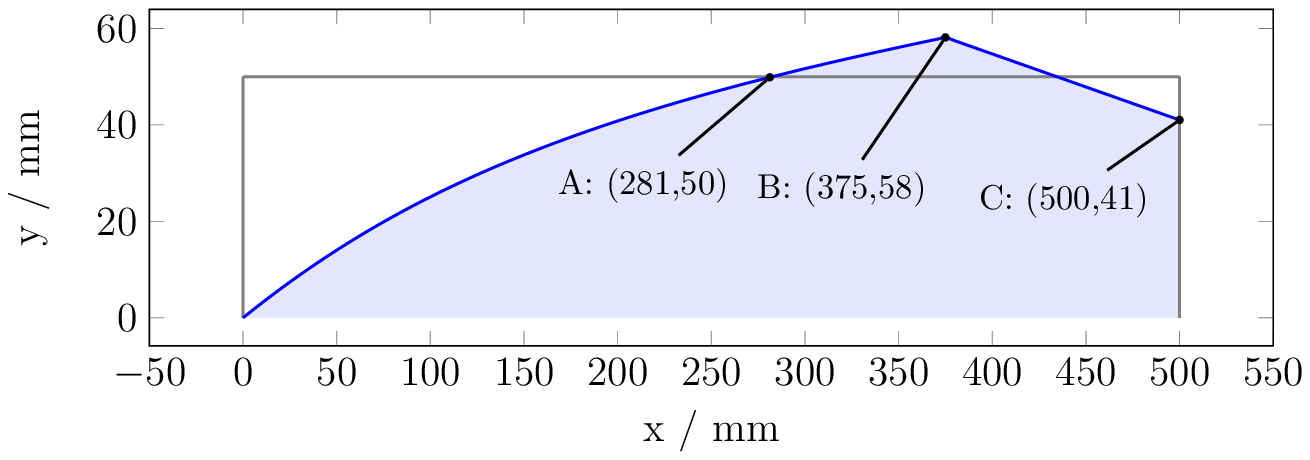}}
	\caption{Optimal satellite geometry for $\alpha\textsubscript{T},g = 0.65$ with shape (c) and 25 $\%$ tail length, achieving $46.3 \: \%$ drag reduction and lifetime extension.}
	\label{fig:optimum-065}
\end{figure}

The best design for further changed satellite surfaces with an even higher fraction of specular particle reflections and less thermal accommodation ($\alpha\textsubscript{T},g = 0.65$) is the optimised profile extended with 3D shape (c) and 25 $\%$ tail length, illustrated in Figure~\ref{fig:optimum-065}.
An improvement of $46.3 \: \%$ in the satellite lifetime (+ 77.3 days) is achieved compared to the reference body.

\begin{figure}
	\centering
	\includegraphics[width=\linewidth]{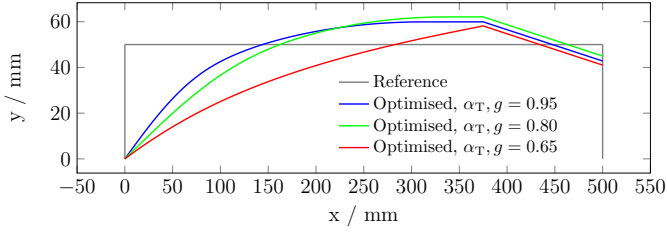}
	\caption{Comparison of the three optimal satellite profiles for the different surface properties.}
	\label{fig:optimum-comparison}
\end{figure}

A comparison of the three optimal satellite profiles for the different surface properties is shown in Figure \ref{fig:optimum-comparison}.
Due to the different volume distributions of the 3D shapes, the integral areas of the blue and green profiles with shape (a) are greater than that of the red profile with shape (c).

To conclude, a short tail of $25 \: \%$ of the total body length is advantageous for a reduced drag design due to a decreased drag coefficient $C\textsubscript{d}$ compared to the optimised geometries without tail variations.
However, longer tail lengths of $50 \: \%$ and $75 \: \%$ of the total body are not preferable since the reference areas of the satellite bodies $A\textsubscript{ref}$ are increased significantly with the tail length due to the constant body volume.

\begin{figure}
	\centering
	\includegraphics[width=0.5\linewidth]{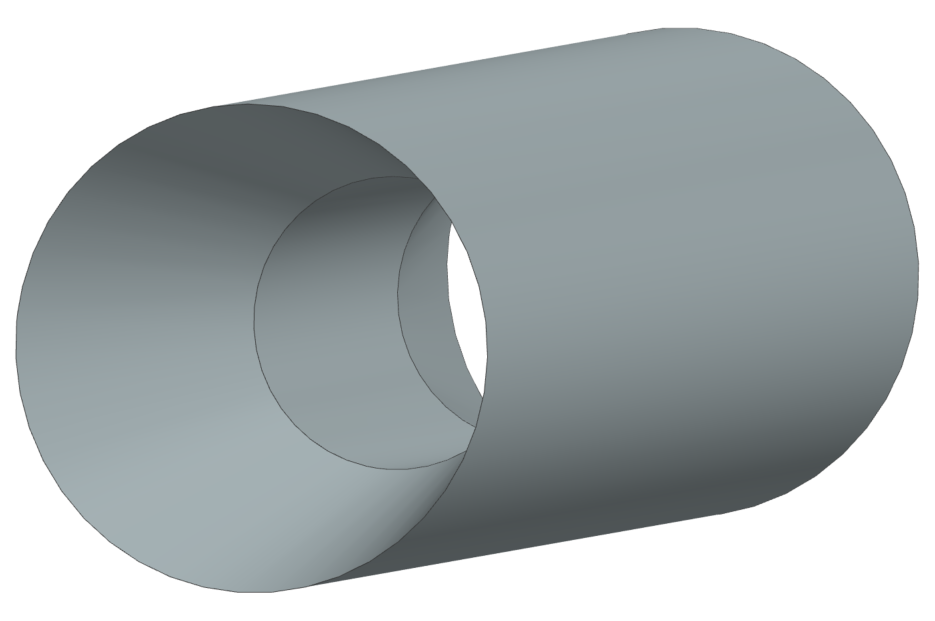}
	\caption{Optimal satellite geometry for $\alpha\textsubscript{T},g = 0.00$: Ring geometry (2).}
	\label{fig:optimum-00}
\end{figure}

Considering the limiting case of fully specular particle reflections ($\alpha\textsubscript{T},g = 0.00$), the presented ring geometry (2), which is depicted in Figure~\ref{fig:optimum-00}, accomplishes the lowest drag and the highest lifetime.
An increase by more than $3300 \: \%$ could be accomplished.

\section{Conclusion}
Due to various advantages, there is a large interest to operate satellites in Very Low Earth Orbits.
However, the main challenge is the drag force acting on the satellites that leads to a reduced lifetime.
Aerodynamic satellite designs including a high ballistic coefficient $\beta$ can achieve a significant drag reduction.

For this, satellite designs have been aerodynamically optimised whereby a maximum lifetime increase has been achieved for bodies including a tail of $25 \: \%$ of the total satellite length.
An increase of $13.0 \: \%$ is possible for state-of-the-art surface properties ($\alpha\textsubscript{T},g = 0.95$) and the presented 3D shape (a), whereas maximal considered material improvements ($\alpha\textsubscript{T},g = 0.65$) yield for a $46.3 \: \%$ increase in lifetime with 3D shape (c), both compared to the cylindrical reference body.
Generally, the application of a long tail length is not advisable since the reference area $A\textsubscript{ref}$ and thus the drag force $F\textsubscript{d}$ increase significantly for satellites with constant volume.
Ring geometries are only preferable for small values of $\alpha\textsubscript{T}$ and $g$, where for the fully specular case, the satellite lifetime is increased by more than $3300 \: \%$ with ring geometry (2).
Here, challenges regarding the manufacturing process and the placement of instruments need to be considered.
In general, the simulations have shown that worthwhile improvements in the satellite's lifetime can be achieved by geometry variations even if the volume is kept constant.

The sensitivity of the drag force acting on the considered bodies towards variations in the angle of attack $\chi$ is limited.
The optimised bodies yield less drag increase as the reference body while the increase in the lift force is improved.

\subsection{Future Work}
Future investigations could comprise an extension of the 2D profile optimisation to a 3D optimisation, which would allow for other designs than the three shapes considered here.
This would require a complete rework of the drag function computation and the specified constraints.
The inclusion of multiple particle reflections as consequences of more complex structures for solar panels as well as booms, e.g. for antennas, would certainly also be of high interest.
Also, an assessment of the effect of the generated lift force on attitude and formation flight control as well as its impact on the lifetime of the satellite for non-symmetrical bodies should be pursued in the future.
An analysis of the attitude stability of the optimised geometries is to be done as well.
Finally, future investigations could consider also the effects of shape optimisation in combination with active countermeasures, e.g. thrusters with on-board propellants and/or air-breathing concepts, in order to minimise overall mass demands. 

\section*{Acknowledgments}
This project has received funding from the European Research Council (ERC) under the European Union’s Horizon 2020 research and innovation programme (grant agreement No. 899981 MEDUSA).

\appendix

\section{Relevant Equations}

\subsection{Sentman Model} \label{app:sentman}
The dimensionless coefficient of the total force component in a particular direction on an area  element assuming free molecular flow according to Sentman~\cite{sentman} is given by:
\begin{equation}
	\begin{split}
		\frac{\mathrm{d} C}{\mathrm{d} A} &= \frac{1}{A\textsubscript{ref}} \Biggl\{ \left[\sigma \cdot \left(\epsilon k + \eta t \right) + \left(2 - \sigma' \right) \cdot \gamma l \right] \Biggr. \\
		&\cdot \left. \left[\gamma \left(1 + \mathrm{erf}(\gamma s) \right) + \frac{1}{s \sqrt{\pi}} \mathrm{e}^{-\gamma^2 s^2} \right] \right. \\
		&+ \left. \frac{\left(2 - \sigma' \right) \cdot l}{2 s^2} \left(1 + \mathrm{erf}(\gamma s) \right) \right. \\
		&+ \Biggl. \frac{\sigma' l}{2} \sqrt{\frac{T\textsubscript{w}}{T\textsubscript{i}}} \left[ \frac{\gamma \sqrt{\pi}}{s} \left(1 + \mathrm{erf}(\gamma s) \right) + \frac{1}{s^2} \mathrm{e}^{-\gamma^2 s^2} \right] \Biggr\}. \label{eq:sentman}
	\end{split}
\end{equation}
The drag force can be calculated from this using Equations~\eqref{eq:F_d} and \eqref{eq:a_d}.

\subsection{Determination of Momentum Accommodation Coefficients}  \label{app:coefficients}
With $E = \frac{1}{2} m v^2$ and $I = mv$~\cite{physik} as well as the assumption of $m = \mathrm{const.}$, the kinetic energy of a particle is proportional to the square of its momentum: $E \propto I^2$.
Thus, the energy accommodation coefficient defined in Equation~\eqref{eq:alpha} can be rewritten as
\begin{equation}
	\alpha\textsubscript{T} = \frac{I\textsubscript{i}^2 - I\textsubscript{r}^2}{I\textsubscript{i}^2 - I\textsubscript{w}^2},
\end{equation}
resulting in a mathematical formulation for the total momentum carried away from the surface by the reflected particles:
\begin{equation}
	I\textsubscript{r} = \sqrt{I\textsubscript{i}^2 - \alpha\textsubscript{T} \left(  I\textsubscript{i}^2 - I\textsubscript{w}^2 \right)}. \label{eq:I_r}
\end{equation}
For complete diffuse reflection, $\tau\textsubscript{r} = 0$ applies~\cite{sentman}.
With the general correlation that the total momentum $I$ can be split into its normal and tangential components,
\begin{equation}
	I^2 = p^2 + \tau^2, \label{eq:I}
\end{equation}
it follows $p\textsubscript{r} = I\textsubscript{r}$.
Considering specular particle reflections, the angle of incidence $\theta\textsubscript{i}$ equals the angle of reflection $\theta\textsubscript{r}$, which is why the intercept theorem is applied:
\begin{equation}
	\left\lvert \frac{p\textsubscript{r}}{p\textsubscript{i}} \right\rvert = \frac{\tau\textsubscript{r}}{\tau\textsubscript{i}} = \left\lvert \frac{I\textsubscript{r}}{I\textsubscript{i}} \right\rvert.
\end{equation}
Subsequently, $\tau\textsubscript{r}$ and $p\textsubscript{r}$ are combined by using the fractions of diffuse and specular reflected particles according to Maxwell:
\begin{align}	
	\tau\textsubscript{r} &= \left( 1 - g \right) \tau\textsubscript{i} \frac{I\textsubscript{r}}{I\textsubscript{i}},\\
	p\textsubscript{r} &= g \cdot I\textsubscript{r} + \left( 1 - g \right) \cdot p\textsubscript{i} \frac{I\textsubscript{r}}{I\textsubscript{i}}.
\end{align}
Inserting these terms into Equations~\eqref{eq:sigma} and \eqref{eq:sigma'}, it follows:
\begin{align}	
	\sigma &= \frac{\tau\textsubscript{i} - \left[ \left( 1 - g \right) \tau\textsubscript{i} \frac{I\textsubscript{r}}{I\textsubscript{i}} \right]}{\tau\textsubscript{i}},\\
	\sigma' &= \frac{p\textsubscript{i} - \left[ g \cdot I\textsubscript{r} + \left( 1 - g \right) \cdot p\textsubscript{i} \frac{I\textsubscript{r}}{I\textsubscript{i}} \right]}{p\textsubscript{i} - p\textsubscript{w}}.
\end{align}
The total momentum of the reflected particles is already defined in Equation \eqref{eq:I_r}.
The total momentum of the incident particles is defined utilising Equation \eqref{eq:I}:
\begin{equation}
	I\textsubscript{i} = \sqrt{p\textsubscript{i}^2 + \tau\textsubscript{i}^2}.
\end{equation}
Additionally, $I\textsubscript{w} = p\textsubscript{w}$, due to $\tau\textsubscript{w} = 0$.
$\tau\textsubscript{i}, p\textsubscript{i}$ and $p\textsubscript{w}$ are determined utilising Equations (21) and (28) from~\cite{sentman}:
\begin{align}
	\tau\textsubscript{i} &= \epsilon \left[ \gamma \left( 1 + \mathrm{erf} \left( \gamma s \right) \right) + \frac{1}{s \sqrt{\pi}} \mathrm{e}^{-\gamma^2 s^2} \right],\\
	p\textsubscript{i} &= \gamma \left[ \gamma \left( 1 + \mathrm{erf} \left( \gamma s \right) \right) + \frac{1}{s \sqrt{\pi}} \mathrm{e}^{-\gamma^2 s^2} \right] \\
	&+ \frac{1}{2 s^2} \left( 1 + \mathrm{erf} \left( \gamma s \right) \right),\\
	p\textsubscript{w} &= \frac{1}{2} \sqrt{\frac{T\textsubscript{w}}{T\textsubscript{i}}} \left[ \frac{\gamma \sqrt{\pi}}{s} \left( 1 + \mathrm{erf} \left( \gamma s \right) \right) + \frac{1}{s^2} \mathrm{e}^{-\gamma^2 s^2} \right].
\end{align}
The reader should note that constant values including the density, the relative velocity, reference area and particle flows are cancelled out and are thus not considered in these equations.

\subsection{Volume Restriction} \label{app:volume}
In the numerical profile optimisation, an equality constraint is implemented for the satellite volume restriction depending on the three-dimensional geometries.

\subsubsection{Volume of Shape (a)}
The volume of a rotated body (shape (a)) is calculated by~\cite{mathematics}:
\begin{equation}
	V_a = \pi \int_{0}^{x\textsubscript{max}} f(x)^2 \,\mathrm{d}x.
\end{equation}
Using the Gaussian quadrature with two support points $y_1 = - \sqrt{\frac{1}{3}}$ and $y_2 = \sqrt{\frac{1}{3}}$~\cite{mathematics}, it follows:
\begin{equation}
	V_a = \sum_{i=2}^{n+1} \left\{ \frac{\pi \Delta x}{2} \cdot \sum_{j=1}^{2} f \left( y_j \frac{\Delta x}{2} + \frac{x_{i-1} + x_i}{2} \right)^2 \right\}.
\end{equation}
This is subsequently rewritten with the definition of the middle points of the sections:
\begin{equation}
	\frac{x_{i-1} + x_i}{2} = x_i + \frac{\Delta x}{2}.
\end{equation}
Thus, it follows:
\begin{align}
	V_a &= \sum_{i=2}^{n+1} \left\{ \frac{\pi \Delta x}{2} \cdot \sum_{j=1}^{2} f \left( x_{i-1} + \frac{\Delta x}{2} \left( 1 + y_j \right) \right)^2 \right\} \nonumber \\
	&= \sum_{i=2}^{n+1} \left\{ \frac{\pi \Delta x}{2} \cdot \sum_{j=1}^{2} \left[ f_{i-1} + \frac{1 + y_i}{2} \cdot \Delta x \cdot \frac{f_i - f_{i-1}}{\Delta x} \right]^2 \right\} \nonumber \\
	&= \sum_{i=2}^{n+1} \left\{ \frac{\pi \Delta x}{2} \cdot \sum_{j=1}^{2} \left[ f_{i-1} \cdot \left( \frac{1 - y_j}{2} \right) + f_i \cdot \left( \frac{1 + y_j}{2} \right) \right]^2 \right\}.	
\end{align}

\subsubsection{Volume of Shape (b)}
The volume of the extruded shape (b) is calculated utilising its profile integral area $A$:
\begin{align}
	A &= \Delta x \cdot
	\left[ {\begin{array}{ccccccc}
			0.5 & 1 & 1 & \cdots & 1 & 1 & 0.5 \\
	\end{array} } \right]
	\textbf{f},\\
	V_b &= 4 \cdot A \cdot f_{n+1}.
\end{align}

\subsubsection{Volume of Shape (c)}
\begin{figure}
	\centering
	\includegraphics[width=0.5\linewidth]{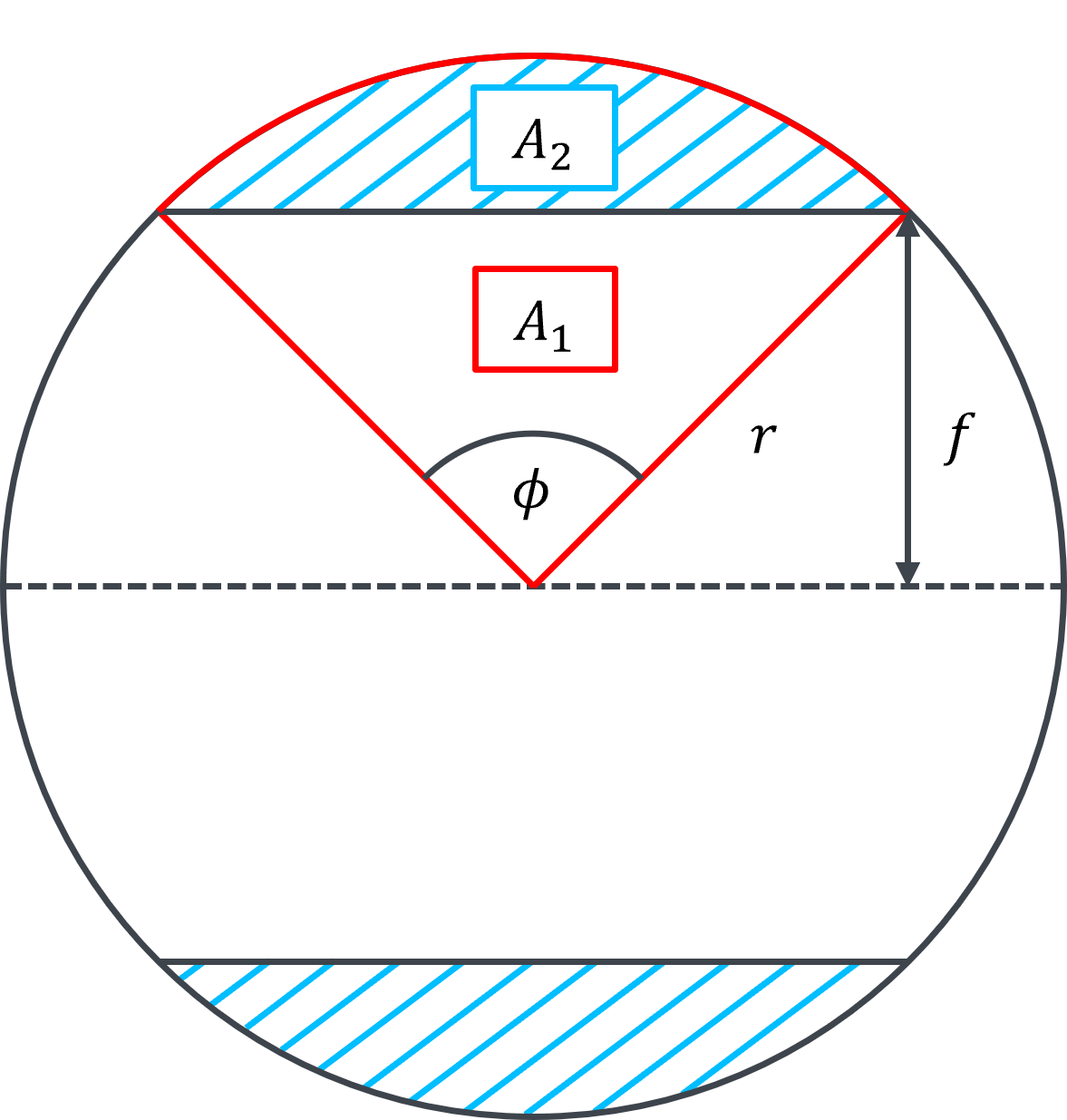}
	\caption{The areas perpendicular to the incident velocity vector for the volume calculation of 3D shape (c).}
	\label{fig:volume-c}
\end{figure}

For shape (c), the volume is assembled from areas perpendicular to the incident velocity vector multiplied with $\Delta x$.
The single areas are calculated out of circular segments depicted in Figure~\ref{fig:volume-c}, where the cross-hatched areas need to be subtracted.
A circle segment with opening angle $\phi$ is calculated by:
\begin{equation}
	A_1 = \frac{\phi}{2} r^2.
\end{equation}
One cross-hatched area $A_2$ is computed by:
\begin{align}
	A_2 &= A_1 - \left( \sin{\frac{\phi}{2}} \cdot r \cdot \cos{\frac{\phi}{2}} \cdot r \right) \nonumber\\
	&= \frac{\phi}{2} r^2 - \frac{1}{2} r^2 \sin{\phi} \nonumber \\
	&= \frac{1}{2} r^2 \left( \phi - \sin{\phi} \right).
\end{align}
Hence, for the area $A$ utilised for volume calculation applies:
\begin{align}
	A &= \pi r^2 - 2 \cdot A_2 \nonumber\\
	&= r^2 \left( \pi - \phi + \sin{\phi} \right). \label{eq:A}
\end{align}
The opening angle $\phi$ depends on the function value $f$:
\begin{equation}
	\phi = \pi - 2 \cdot \arcsin\left( \frac{f}{r} \right). \label{eq:A_beta}
\end{equation}
The volume is calculated as:
\begin{equation}
	V_c = \int_{0}^{x\textsubscript{max}} A \,\mathrm{d} x.
\end{equation}
Approximating every interval by the mid-point, $f$ becomes $\frac{1}{2} \left( f_i + f_{i-1} \right)$ and the volume is transformed to a sum:
\begin{align}
	V_c &= \sum_{i=2}^{n+1} A_i \Delta x \nonumber\\
	&= \sum_{i=2}^{n+1} \left[ 2 \arcsin \left( \frac{f_i + f_{i-1}}{2 r} \right) \right. \nonumber\\
	&+ \left. \sin \left( \pi - 2 \arcsin \left( \frac{f_i + f_{i-1}}{2 r} \right) \right) \right] r^2 \Delta x.
\end{align}
Equations~\eqref{eq:A} and \eqref{eq:A_beta} are included and $r = f_{n+1}$ applies.

 \bibliographystyle{elsarticle-num} 
 \bibliography{cas-refs}





\end{document}